# Proceedings Scholar Metrics:
# H Index of proceedings on Computer Science, Electrical & Electronic Engineering, and Communications according to Google Scholar Metrics (2010-2014)

Alberto Martín-Martín[1], Juan Manuel Ayllón[1], Enrique Orduña-Malea[2], Emilio Delgado López-Cózar[1]

[1]EC3: Evaluación de la Ciencia y de la Comunicación Científica. Universidad de Granada
[2]EC3: Evaluación de la Ciencia y de la Comunicación Científica, Universidad Politécnica de Valencia (Spain)

**ABSTRACT**

The objective of this report is to present a list of proceedings (conferences, workshops, symposia, meetings) in the areas of Computer Science, Electrical & Electronic Engineering, and Communications covered by Google Scholar Metrics and ranked according to their h-index. Google Scholar Metrics only displays publications that have published at least 100 papers and have received at least one citation in the last five years (2010-2014). The searches were conducted between the 8th and 10th of December, 2015. A total of 1501 proceedings have been identified.

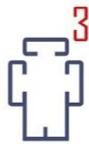

# INTRODUCTION

**PROCEEDINGS SCHOLAR METRICS** is a ranking that displays proceedings (conferences, workshops, symposia, meetings) indexed in Google Scholar Metrics (GSM) on the areas of Computer Science, Electrical & Electronic Engineering, and Communications for the period 2010-2014.

It is a well-known fact that conference proceedings play a major role as a means of scientific communication in all areas concerning Computer Engineering, Electronics, and Communications. The rapid rate at which knowledge is generated in these fields required the creation of a more dynamic system to communicate and publish research results. Conferences have historically fulfilled this role.

Therefore, it is not surprising that these publications take up an important place in researchers' curriculum and that it is an object of special consideration in researchers' performance evaluation programmes.

The various databases that have been traditionally used for evaluative purposes (Web of Science, Scopus) cover these types of publications with varying degrees of exhaustiveness, but never optimally. In fact, this has been one of the reasons why there have already been various attempts at creating classifications that identify and rank these types of publications.

The creation of Google Scholar in 2004, an academic search engine dedicated to crawling the Web in search of scientific literature, revolutionized the world of scientific information search systems. Particularly, it had very positive effect on those disciplines where publication habits were not limited to publishing in scientific journals, i.e. Engineering, Computer Science, Electrical & Electronic Engineering, and Communications.

Publications in these disciplines, for which the Internet is the natural environment, are splendidly represented on the Web. That is why Google Scholar is especially useful in these fields.

The development of Google Scholar Metrics, launched on April 2012 with the goal of providing a ranking of scientific publications indexed on Google Scholar (journals, proceedings, repositories), provided that they had published at least 100 papers and received at least one citation in the last five years, has been a crucial step towards knowing the impact of conferences, which are so important in these areas.

In GSM, publications rankings are sorted by impact (h index) and they can be browsed by languages, each of these showing the 100 publications with the greatest impact. For publications written in English, this tool allows user to browse publications by subject categories and subcategories. Computer Science, Engineering and Communications can be found on the "Engineering & Computer Science Category", which comprises 58 subcategories. For each of these subcategories, only the top 20 publications with a highest H Index are displayed. There are a total of 190 unique conference proceedings across these subcategories. Categories and subcategories are not available for the other eight languages present in GSM (Chinese, Portuguese, German, Spanish, French, Italian, Japanese and Duth). In order to find other publications not displayed in the language and category league lists, there is a search box at our disposal. However, any given query will only display 20 results.

Therefore, the system interface does not allow us to effectively determine which and how many conference proceedings GSM has indexed. In order to overcome this limitation, the objective of **PROCEEDINGS SCHOLAR METRICS** is to compile an inventory of all the conferences present in GSM concerning these fields of knowledge, and after that, rank them according to their scientific impact, as measured by the H index.

## MATERIAL AND METHODS

### Subject areas covered

Proceedings concerning Computer Science (theoretical, information theory, artificial intelligence, evolutionary computation, fuzzy systems, human computer interaction, computer vision & pattern recognition, computer hardware design, computing systems, signal processing, computer networks & wireless communication, robotics, automation & control theory, software systems, computer security & cryptography, computer graphics, databases & information systems, data mining & analysis, multimedia, bioinformatics & computational biology, biomedical technology, medical informatics, computational linguistics, education technology), Electrical & Electronic Engineering and Communications (telecommunications, remote sensing, antennas, radar, microware).

### Search strategy

In order to identify the proceedings we followed two different strategies:

1. We collected all the conferences displayed in the subcategories concerning Computer Science, Electrical & Electronic Engineering, and Communications.
2. We carried out various searches using descriptive and pertinent keywords in order to locate the rest of relevant conferences. These searches took place on the second week of December, 2015.

The results of each search were downloaded (including the name of the conference, the h5-index, and the h5-median) and duplicates were removed. A manual check was carried out in order to filter out any irrelevant entries (journals, repositories, and conferences outside the scope of our study). A total of 1501 conference proceedings were identified.

### Criteria for the inclusion of Google Scholar Metrics proceedings

Proceedings that published at least 100 papers and received at least one citation in the last five years (2010-2014)

### Sorting criteria and fields displayed

The proceedings are sorted by their H Index. In case of a tie, the discriminate value is the h5-median (the median number of citations for the articles that make up its h-index)

The information displayed for each conference is:
- H5-Index
- H5-median
- Quartile: quartile position of the conference.

# H Index of proceedings on Computer Science, Electrical & Electronic Engineering, and Communications according to Google Scholar Metrics (2010-2014)

| Rank | Quartil | Proceedings | h-index | h5-median |
|---|---|---|---|---|
| 1 | Q1 | IEEE Conference on Computer Vision and Pattern Recognition, CVPR | 128 | 203 |
| 2 | Q1 | Proceedings of the IEEE | 87 | 164 |
| 3 | Q1 | Computer Human Interaction (CHI) | 84 | 124 |
| 4 | Q1 | European Conference on Computer Vision | 79 | 133 |
| 5 | Q1 | Annual Joint Conference of the IEEE Computer and Communications Societies (INFOCOM) | 78 | 106 |
| 6 | Q1 | International World Wide Web Conferences (WWW) | 75 | 125 |
| 7 | Q1 | Neural Information Processing Systems (NIPS) | 72 | 101 |
| 8 | Q1 | International Conference on Machine Learning (ICML) | 70 | 115 |
| 9 | Q1 | IEEE International Conference on Computer Vision, ICCV | 68 | 105 |
| 10 | Q1 | International Conference on Very Large Databases | 68 | 92 |
| 11 | Q1 | Meeting of the Association for Computational Linguistics (ACL) | 65 | 99 |
| 12 | Q1 | ACM SIGCOMM Conference | 64 | 118 |
| 13 | Q1 | ACM SIGKDD International Conference on Knowledge discovery and data mining | 64 | 100 |
| 14 | Q1 | IEEE International Conference on Robotics and Automation | 64 | 94 |
| 15 | Q1 | ACM SIGMOD International Conference on Management of Data | 62 | 100 |
| 16 | Q1 | IEEE International Solid-State Circuits Conference | 59 | 84 |
| 17 | Q1 | IEEE Symposium on Security and Privacy | 58 | 103 |
| 18 | Q1 | ACM International Conference on Web Search and Data Mining | 58 | 90 |
| 19 | Q1 | ACM Symposium on Computer and Communications Security | 58 | 87 |
| 20 | Q1 | ACM Symposium on Theory of Computing | 57 | 80 |
| 21 | Q1 | Conference on Empirical Methods in Natural Language Processing (EMNLP) | 56 | 81 |
| 22 | Q1 | International Conference on Software Engineering | 56 | 77 |
| 23 | Q1 | International Conference on Weblogs and Social Media | 55 | 105 |
| 24 | Q1 | Symposium on Networked Systems: Design and Implementation (NSDI) | 55 | 98 |
| 25 | Q1 | IEEE International Conference on Acoustics, Speech and Signal Processing (ICASSP) | 54 | 73 |
| 26 | Q1 | ACM SIGIR Conference on Research and Development in Information Retrieval | 52 | 69 |
| 27 | Q1 | International Conference on Architectural Support for Programming Languages and Operating Systems (ASPLOS) | 51 | 92 |
| 28 | Q1 | International Symposium on Computer Architecture (ISCA) | 51 | 80 |
| 29 | Q1 | USENIX Conference on Security | 50 | 82 |
| 30 | Q1 | International Conference on Data Engineering Workshops | 50 | 67 |
| 31 | Q1 | International Conference on Mobile systems, applications, and services | 49 | 111 |
| 32 | Q1 | Conference on Advances in cryptology | 49 | 79 |
| 33 | Q1 | Supercomputing (SC) | 49 | 71 |
| 34 | Q1 | Annual International Conference on Theory and Applications of Cryptographic Techniques (EUROCRYPT) | 48 | 93 |
| 35 | Q1 | AAAI Conference on Artificial Intelligence | 48 | 65 |
| 36 | Q1 | IEEE International Conference on Cloud Computing (CLOUD) | 47 | 73 |
| 37 | Q1 | IEEE Symposium on Foundations of Computer Science (FOCS) | 47 | 67 |
| 38 | Q1 | ACM SIGPLAN-SIGACT Symposium on Principles of Programming Languages (POPL) | 47 | 66 |
| 39 | Q1 | ACM Computing Surveys (CSUR) | 46 | 105 |
| 40 | Q1 | Internet Measurement Conference | 46 | 83 |
| 41 | Q1 | IEEE International Symposium on Information Theory | 46 | 74 |
| 42 | Q1 | Conference on Computer Supported Cooperative Work (CSCW) | 46 | 72 |
| 43 | Q1 | IEEE GLOBECOM Workshops | 46 | 71 |
| 44 | Q1 | Health technology assessment (Winchester, England) | 46 | 70 |
| 45 | Q1 | IEEE International Conference on Communications | 46 | 62 |
| 46 | Q1 | ACM SIAM Symposium on Discrete Algorithms | 46 | 57 |
| 47 | Q1 | ACM International Conference on Multimedia | 45 | 74 |
| 48 | Q1 | Network and Distributed System Security Symposium (NDSS) | 44 | 76 |
| 49 | Q1 | SIGPLAN Conference on Programming Language Design and Implementation (PLDI) | 44 | 70 |
| 50 | Q1 | IEEE International Electron Devices Meeting, IEDM | 44 | 65 |
| 51 | Q1 | IEEE/RSJ International Conference on Intelligent Robots and Systems | 44 | 64 |
| 52 | Q1 | IEEE Conference on Decision and Control | 43 | 69 |
| 53 | Q1 | IEEE International Symposium on Parallel & Distributed Processing | 43 | 64 |
| 54 | Q1 | American Control Conference | 43 | 55 |
| 55 | Q1 | International Joint Conference on Artificial Intelligence (IJCAI) | 43 | 53 |
| 56 | Q1 | ACM european Conference on Computer Systems | 42 | 87 |
| 57 | Q1 | Annual International Conference on Mobile computing and networking | 42 | 83 |
| 58 | Q1 | Quantum Electronics and Laser Science Conference | 42 | 74 |
| 59 | Q1 | Design Automation Conference (DAC) | 42 | 61 |
| 60 | Q1 | ACM Symposium on User Interface Software and Technology | 41 | 66 |
| 61 | Q1 | ACM International Conference on Information and Knowledge Management | 41 | 65 |
| 62 | Q1 | Design, Automation and Test in Europe Conference and Exhibition (DATE) | 41 | 60 |
| 63 | Q1 | ACM Symposium on Cloud Computing | 40 | 84 |
| 64 | Q1 | Handheld and Ubiquitous Computing (HUC) | 40 | 71 |
| 65 | Q1 | IEEE International Symposium on High Performance Computer Architecture | 40 | 69 |
| 66 | Q1 | International Conference on The Semantic Web | 40 | 59 |
| 67 | Q1 | International Conference on Autonomous Agents and Multiagent Systems | 40 | 58 |
| 68 | Q1 | International Conference on Spoken Language Processing (INTERSPEECH) | 39 | 70 |
| 69 | Q1 | IEEE/ACM International Symposium on Microarchitecture | 39 | 59 |
| 70 | Q1 | Hawaii International Conference on System Sciences | 39 | 57 |
| 71 | Q1 | IEEE International Conference on Data Mining (ICDM) | 39 | 53 |
| 72 | Q1 | Conference on File and Storage Technologies (FAST) | 38 | 75 |
| 73 | Q1 | International Conference on Language Resources and Evaluation | 38 | 64 |
| 74 | Q1 | IEEE Vehicular Technology Conference, VTC | 38 | 62 |
| 75 | Q1 | IEEE International Symposium on Cluster Computing and the Grid | 38 | 59 |
| 75 | Q1 | International Conference on Computational Linguistics (COLING) | 38 | 59 |
| 77 | Q1 | ACM Conference on Recommender Systems | 38 | 55 |
| 77 | Q1 | ACM SIGMETRICS Performance Evaluation Review | 38 | 55 |
| 79 | Q1 | International Conference on Distributed Computing Systems, ICDCS | 38 | 54 |
| 80 | Q1 | International Conference on Computer Aided Verification (CAV) | 38 | 51 |
| 81 | Q1 | IEEE International Conference on Image Processing (ICIP) | 38 | 46 |
| 82 | Q1 | British Machine Vision Conference (BMVC) | 37 | 68 |
| 83 | Q1 | IEEE Applied Power Electronics Conference and Exposition | 37 | 52 |
| 83 | Q1 | USENIX Annual Technical Conference | 36 | 57 |
| 85 | Q1 | International Conference on The Theory and Application of Cryptology and Information Security (ASIACRYPT) | 35 | 58 |
| 86 | Q1 | ACM SIGSOFT International Symposium on Foundations of Software Engineering | 35 | 56 |
| 87 | Q1 | IEEE International Conference on Cloud Computing Technology and Science (CloudCom) | 35 | 55 |
| 88 | Q1 | Extended Semantic Web Conference | 35 | 53 |
| 89 | Q1 | Allerton Conference on Communication, Control, and Computing | 35 | 50 |
| 90 | Q1 | ACM Conference on Electronic Commerce | 35 | 47 |
| 90 | Q1 | SIAM International Conference on Data Mining | 35 | 47 |
| 92 | Q1 | ACM Symposium on Applied Computing | 35 | 42 |
| 93 | Q1 | International Conference on Parallel Architectures and Compilation Techniques (PACT) | 34 | 56 |
| 94 | Q1 | ACM SIGPLAN Symposium on Principles & Practice of Parallel Programming (PPOPP) | 34 | 54 |
| 95 | Q1 | Workshop on Cryptographic Hardware and Embedded Systems (CHES) | 34 | 51 |
| 96 | Q1 | International Conference on Pattern Recognition | 34 | 50 |
| 97 | Q1 | Conference on Object-Oriented Programming Systems, Languages, and Applications (OOPSLA) | 34 | 48 |
| 98 | Q1 | International Conference on Extending Database Technology (EDBT) | 34 | 46 |
| 99 | Q1 | IEEE Wireless Communications & Networking Conference | 34 | 44 |
| 100 | Q1 | International Symposium on Information Processing in Sensor Networks | 33 | 65 |
| 101 | Q1 | Conference on Emerging Network Experiment and Technology (CoNEXT) | 33 | 57 |
| 102 | Q1 | ACM Conference on Embedded Networked Sensor Systems | 33 | 55 |
| 103 | Q1 | IEEE Conference on Computer Vision and Pattern Recognition Workshops (CVPRW) | 33 | 53 |
| 104 | Q1 | IEEE/ACM International Conference on Automated Software Engineering (ASE) | 33 | 38 |
| 105 | Q1 | ACM/IEEE International Conference on Human Robot Interaction | 32 | 52 |
| 106 | Q1 | ACM International Symposium on High Performance Distributed Computing | 32 | 51 |
| 107 | Q1 | International Conference on Social Computing | 32 | 50 |
| 108 | Q1 | Conference in Uncertainty in Artificial Intelligence | 32 | 47 |
| 109 | Q1 | International Conference on Tools and Algorithms for the Construction and Analysis of Systems (TACAS) | 32 | 46 |
| 110 | Q1 | Conference on Genetic and Evolutionary Computation | 32 | 42 |
| 111 | Q1 | International Conference on Practice and Theory in Public Key Cryptography | 31 | 50 |
| 112 | Q1 | International Society for Music Information Retrieval Conference | 31 | 46 |
| 113 | Q1 | Computer Security Applications Conference | 31 | 45 |
| 113 | Q1 | International Symposium on Software Testing and Analysis | 31 | 45 |
| 115 | Q1 | IEEE Congress on Evolutionary Computation | 31 | 43 |
| 116 | Q1 | International Colloquium on Automata, Languages and Programming (ICALP) | 31 | 41 |
| 117 | Q1 | IFAC World Congress | 31 | 37 |
| 118 | Q1 | ACM Workshop on Hot Topics in Networks | 30 | 48 |
| 119 | Q1 | European Conference and Exposition on Optical Communications | 30 | 42 |
| 120 | Q1 | ACM SIGMOD-SIGACT-SIGART Symposium on Principles of Database Systems | 30 | 40 |
| 120 | Q1 | Asian Conference on Computer Vision | 30 | 40 |
| 120 | Q1 | IEEE International Conference on Automatic Face & Gesture Recognition and Workshops | 30 | 40 |
| 123 | Q1 | IEEE Information Theory Workshop | 30 | 39 |
| 124 | Q1 | International Conference on Financial Cryptography and Data Security | 29 | 53 |
| 125 | Q1 | IEEE Symposium on VLSI Technology | 29 | 51 |
| 126 | Q1 | IEEE International Symposium on Circuits and Systems | 29 | 43 |
| 127 | Q1 | IEEE International Conference on Web Services | 29 | 37 |
| 128 | Q1 | IEEE International Conference on Computer Vision Workshops (ICCV Workshops) | 28 | 52 |
| 129 | Q1 | IEEE Symposium on New Frontiers in Dynamic Spectrum Access Networks | 28 | 49 |
| 130 | Q1 | European Conference on Optical Communications | 28 | 41 |
| 130 | Q1 | International Joint Conference on Neural Networks | 28 | 41 |
| 132 | Q1 | European Conference on Machine learning and knowledge discovery in databases | 28 | 40 |
| 132 | Q1 | IEEE Conference of Industrial Electronics | 28 | 40 |
| 132 | Q1 | International Power Electronics Conference (IPEC) | 28 | 40 |
| 135 | Q1 | IEEE/MTT-S International Microwave Symposium | 28 | 39 |
| 136 | Q1 | ACM SIGSPATIAL International Conference on Advances in Geographic Information Systems | 28 | 37 |
| 136 | Q1 | IEEE Symposium on Logic in Computer Science | 28 | 37 |
| 138 | Q1 | Principles of Knowledge Representation and Reasoning (KR) | 28 | 36 |
| 139 | Q1 | ACM Symposium on Information, Computer and Communications Security | 27 | 59 |
| 140 | Q1 | Workshop of Cross-Language Evaluation Forum | 27 | 41 |
| 141 | Q1 | European Conference on Parallel Processing (Euro-Par) | 27 | 40 |
| 141 | Q1 | IEEE International Conference on Pervasive Computing and Communications (PerCom) | 27 | 40 |
| 143 | Q1 | Asilomar Conference on Signals, Systems and Computers | 27 | 39 |
| 143 | Q1 | IEEE International Conference on Software Testing, Verification and Validation Workshops (ICSTW) | 27 | 39 |
| 145 | Q1 | Euromicro Conference on Real-Time Systems | 27 | 38 |
| 145 | Q1 | IEEE International Conference on Software Maintenance | 27 | 38 |
| 145 | Q1 | International Symposium on Code Generation and Optimization | 27 | 38 |
| 148 | Q1 | International Conference on Topics in Cryptology | 27 | 37 |
| 149 | Q1 | IEEE International Conference on Pervasive Computing and Communications Workshops (PERCOM Workshops) | 27 | 36 |
| 149 | Q1 | IEEE/IFIP International Conference on Dependable Systems and Networks | 27 | 36 |
| 151 | Q1 | Asia and South Pacific Design Automation Conference (ASP-DAC) | 27 | 35 |
| 152 | Q1 | Electronic Components and Technology Conference, ECTC | 27 | 32 |
| 153 | Q1 | International Conference on Supercomputing (ICS) | 26 | 52 |
| 154 | Q1 | European Conference on Advances in Information Retrieval | 26 | 44 |
| 154 | Q1 | IEEE Real-Time Systems Symposium (RTSS) | 26 | 44 |
| 156 | Q1 | ACM International Symposium on Mobile Ad Hoc Networking and Computing | 26 | 42 |
| 156 | Q1 | IEEE Computer Security Foundations Symposium | 26 | 42 |
| 158 | Q1 | Fast Software Encryption (FSE) | 26 | 41 |
| 159 | Q1 | ACM Symposium on Parallelism in Algorithms and Architectures (SPAA) | 26 | 40 |
| 160 | Q1 | IEEE/ACM International Conference on Computer-Aided Design (ICCAD) | 26 | 39 |
| 161 | Q1 | European Symposium on Programming (ESOP) | 26 | 38 |
| 161 | Q1 | IEEE Real-Time and Embedded Technology and Applications Symposium | 26 | 38 |
| 163 | Q1 | International Conference on Functional Programming (ICFP) | 26 | 36 |
| 163 | Q1 | International Software Product Line Conference | 26 | 36 |
| 163 | Q1 | Technical Symposium on Computer Science Education | 26 | 36 |
| 166 | Q1 | IEEE Conference on Computational Intelligence and Games | 26 | 35 |
| 167 | Q1 | IEEE International Symposium on Personal, Indoor and Mobile Radio Communications | 26 | 33 |
| 168 | Q1 | USENIX Conference on Hot topics in cloud computing | 25 | 55 |
| 169 | Q1 | European Conference on Object-oriented Programming (ECOOP) | 25 | 45 |
| 170 | Q1 | Conference on Innovative Data system Research (CIDR) | 25 | 39 |
| 171 | Q1 | IEEE International Conference on Advanced Information Networking and Applications | 25 | 34 |
| 171 | Q1 | International Conference on Automated Planning and Scheduling (ICAPS) | 25 | 34 |
| 173 | Q1 | IEEE International Conference on Smart Grid Communications (SmartGridComm) | 25 | 32 |
| 173 | Q1 | International Conference on Biometrics | 25 | 32 |
| 175 | Q1 | ACM Symposium on Principles of Distributed Computing (PODC) | 25 | 31 |
| 176 | Q1 | International Conference on Learning Analytics and Knowledge | 24 | 51 |
| 177 | Q1 | International Conference on Autonomic computing | 24 | 50 |
| 178 | Q1 | ACM Multimedia Systems Conference (MMSys) | 24 | 48 |
| 179 | Q1 | Conference on Information Sciences and Systems | 24 | 43 |
| 180 | Q1 | IEEE International Reliability Physics Symposium (IRPS) | 24 | 41 |
| 180 | Q1 | IEEE Symposium on VLSI Circuits | 24 | 41 |
| 180 | Q1 | International Conference on Document Analysis and Recognition | 24 | 41 |
| 183 | Q1 | European Conference on Research in Computer Security | 24 | 39 |
| 183 | Q1 | International Conference on Tangible, Embedded and Embodied Interaction | 24 | 39 |
| 185 | Q1 | IEEE International Conference on Network Protocols | 24 | 38 |
| 186 | Q1 | Conference on Computational Natural Language Learning | 24 | 36 |
| 186 | Q1 | IEEE Consumer Communications and Networking Conference | 24 | 36 |
| 188 | Q1 | ACM International Conference on Interactive Tabletops and Surfaces (ITS) | 24 | 35 |
| 188 | Q1 | IEEE International Symposium on Industrial Electronics | 24 | 35 |
| 188 | Q1 | IEEE-RAS International Conference on Humanoid Robots (Humanoids) | 24 | 35 |
| 188 | Q1 | International Conference on Intelligent User Interfaces (IUI) | 24 | 35 |
| 188 | Q1 | Symposium on Interactive 3D Graphics (SI3D) | 24 | 35 |
| 193 | Q1 | European Conference on Algorithms | 24 | 34 |
| 193 | Q1 | International Conference on Information Systems (ICIS) | 24 | 34 |
| 195 | Q1 | IEEE International Conference on Services Computing | 24 | 33 |
| 195 | Q1 | Information Theory and Applications Workshop | 24 | 33 |
| 197 | Q1 | ACM/IEEE International Symposium on Networks-on-Chip | 24 | 32 |
| 197 | Q1 | Hybrid Systems: Computation and Control (HSCC) | 24 | 32 |
| 197 | Q1 | IEEE Intelligent Transportation Systems Conference | 24 | 32 |
| 197 | Q1 | IEEE International Conference on Data Mining Workshops (ICDMW) | 24 | 32 |
| 197 | Q1 | Symposium on Computational Geometry | 24 | 32 |
| 202 | Q1 | IEEE International Geoscience and Remote Sensing Symposium | 24 | 31 |
| 203 | Q1 | IEEE International Conference on Multimedia and Expo | 24 | 28 |
| 204 | Q1 | International Quantum Electronics Conference | 23 | 47 |
| 205 | Q1 | Symposium on Field Programmable Gate Arrays (FPGA) | 23 | 43 |
| 206 | Q1 | International Workshop on Semantic Evaluation | 23 | 41 |
| 206 | Q1 | Recent Advances in Intrusion Detection (RAID) | 23 | 41 |
| 208 | Q1 | Computational Learning Theory (COLT) | 23 | 40 |
| 209 | Q1 | Conference on Designing interactive systems | 23 | 35 |
| 209 | Q1 | IEEE International Conference on Advanced Video and Signal-Based Surveillance (AVSS) | 23 | 35 |
| 211 | Q1 | Americas Conference on Information Systems (AMCIS) | 23 | 34 |
| 212 | Q1 | IEEE Communications Society Conference on Sensor, Mesh and Ad Hoc Communications and Networks (SECON) | 23 | 33 |
| 212 | Q1 | IEEE Pacific Visualization Symposium | 23 | 33 |
| 214 | Q1 | European Conference on Artificial Intelligence (ECAI) | 23 | 32 |
| 214 | Q1 | IEEE International Conference on Cluster Computing | 23 | 32 |
| 214 | Q1 | International Conference on Advances in Social Networks Analysis and Mining | 23 | 32 |
| 217 | Q1 | Symposium on Theoretical Aspects of Computer Science (STACS) | 23 | 30 |
| 218 | Q1 | European Conference on Information Systems | 23 | 29 |
| 218 | Q1 | International Conference on Fundamental Approaches to Software Engineering | 23 | 29 |
| 220 | Q1 | European Conference on Software Maintenance and Reengineering | 23 | 28 |
| 221 | Q1 | IEEE Radio Frequency Integrated Circuits Symposium | 23 | 27 |
| 222 | Q1 | International IFIP TC 6 Conference on Networking | 22 | 42 |
| 223 | Q1 | ACM SIGGRAPH/Eurographics Symposium on Computer Animation | 22 | 37 |
| 224 | Q1 | International Conference on Multimodal Interfaces (ICMI) | 22 | 36 |
| 225 | Q1 | ACM/IEEE International Symposium on Low Power Electronics and Design | 22 | 35 |
| 226 | Q1 | IEEE Symposium on Mass Storage Systems and Technologies | 22 | 34 |
| 227 | Q1 | IEEE Symposium on Visual Analytics Science and Technology | 22 | 33 |
| 227 | Q1 | International Conference on Mobile Data Management | 22 | 33 |
| 229 | Q1 | IEEE Custom Integrated Circuits Conference, CICC | 22 | 32 |
| 229 | Q1 | IEEE International Conference on Power Electronics and ECCE Asia (ICPE & ECCE) | 22 | 32 |
| 229 | Q1 | International Conference on Concurrency Theory (CONCUR) | 22 | 32 |
| 229 | Q1 | International Symposium on Empirical Software Engineering and Measurement, ESEM | 22 | 32 |
| 233 | Q1 | IEEE International Conference on Program Comprehension | 22 | 31 |
| 233 | Q1 | IEEE International Symposium on Mixed and Augmented Reality | 22 | 31 |
| 233 | Q1 | IEEE Virtual Reality Conference | 22 | 31 |
| 233 | Q1 | International Conference on Automated Deduction (CADE) | 22 | 31 |
| 233 | Q1 | Verification, Model Checking and Abstract Interpretation (VMCAI) | 22 | 31 |
| 238 | Q1 | ACM Conference on Hypertext and Hypermedia | 22 | 30 |
| 238 | Q1 | IEEE International Symposium on Performance Analysis of Systems and Software | 22 | 30 |
| 238 | Q1 | International Conference on Consumer Electronics | 22 | 30 |
| 238 | Q1 | Proceedings of the ESSCIRC | 22 | 30 |
| 242 | Q1 | IEEE International Conference on Systems, Man and Cybernetics | 22 | 27 |
| 242 | Q1 | IEEE International Symposium on a World of Wireless, Mobile and Multimedia Networks | 22 | 27 |
| 244 | Q1 | ACM International Conference on Multimedia Retrieval | 22 | 26 |
| 245 | Q1 | IEEE International Symposium on Robot and Human Interactive Communication | 22 | 26 |
| 246 | Q1 | AMIA Symposium | 22 | 25 |
| 247 | Q1 | Computer Supported Cooperative Work (CSCW) | 21 | 39 |
| 248 | Q1 | Formal Methods in Computer-Aided Design (FMCAD) | 21 | 38 |
| 248 | Q1 | IEEE International Symposium on Modeling, Analysis & Simulation of Computer and Telecommunication Systems, MASCOTS | 21 | 38 |
| 250 | Q1 | IEEE/ACM/IFIP International Conference on Hardware/Software Codesign and System Aynthesis | 21 | 37 |
| 251 | Q1 | International Conference on Parallel Processing (ICPP) | 21 | 36 |
| 251 | Q1 | International Conference on Selected areas in cryptography | 21 | 36 |
| 253 | Q1 | International Conference on Computer Communications and Networks (ICCCN) | 21 | 35 |
| 254 | Q1 | ACM Symposium on Access Control Models and Technologies (SACMAT) | 21 | 34 |
| 254 | Q1 | Conference of the European Chapter of the Association for Computational Linguistics (EACL) | 21 | 34 |
| 254 | Q1 | Pacific Symposium on Biocomputing | 21 | 34 |
| 257 | Q1 | IEEE Conference on Computational Complexity | 21 | 32 |
| 257 | Q1 | International ACM/IEEE Joint Conference on Digital Libraries | 21 | 32 |
| 257 | Q1 | Workshop on Applications of Computer Vision (WACV) | 21 | 32 |
| 260 | Q1 | Business Process Management | 21 | 30 |
| 260 | Q1 | IEEE International Conference on Computer and Information Technology (CIT) | 21 | 30 |
| 260 | Q1 | Theory and Applications of Satisfiability Testing | 21 | 30 |
| 263 | Q1 | IEEE International Conference on Fuzzy Systems (FUZZ) | 21 | 29 |
| 263 | Q1 | International Conference on Experimental Algorithms | 21 | 29 |
| 263 | Q1 | International Symposium on Software Reliability Engineering | 21 | 29 |
| 266 | Q1 | IEEE Symposium on Field-Programmable Custom Computing Machines | 21 | 28 |
| 266 | Q1 | International Conference on Advanced Information Networking and Applications-Workshops, AINAW | 21 | 28 |
| 268 | Q1 | Intelligent Virtual Agents | 21 | 27 |
| 268 | Q1 | International Conference on Parallel Processing Workshops | 21 | 27 |
| 268 | Q1 | International Wireless Communications and Mobile Computing Conference (IWCMC) | 21 | 27 |
| 271 | Q1 | IEEE Annual Computer Software and Applications Conference Workshops (COMPSACW) | 21 | 26 |
| 271 | Q1 | International ACM Conference on Assistive Technologies (Assets) | 21 | 26 |
| 271 | Q1 | International Conference on Applications of Evolutionary Computation | 21 | 26 |
| 271 | Q1 | International Conference on Field Programmable Logic and Applications | 21 | 26 |
| 271 | Q1 | Static Analysis (WSA/SAS) | 21 | 26 |
| 276 | Q1 | International Conference on Aspect-Oriented Software Development (AOSD) | 20 | 40 |
| 277 | Q1 | Picture Coding Symposium (PCS) | 20 | 36 |
| 278 | Q1 | International Conference on Advanced information systems engineering | 20 | 35 |
| 279 | Q1 | ACM Conference on Data and application security and privacy | 20 | 33 |
| 279 | Q1 | ACM International Conference on Embedded Software | 20 | 33 |
| 281 | Q1 | International Communication Systems and Networks and Workshops COMSNETS | 20 | 32 |
| 282 | Q1 | IEEE International Conference on Mobile Adhoc and Sensor Systems (MASS) | 20 | 31 |
| 282 | Q1 | International Conference on Pervasive Computing Technologies for Healthcare | 20 | 31 |
| 284 | Q1 | European Wireless Conference (EW) | 20 | 30 |
| 284 | Q1 | International Conference on Research in Computational Molecular Biology | 20 | 30 |
| 284 | Q1 | Pacific-Asia Conference on Advances in Knowledge Discovery and Data Mining | 20 | 30 |
| 287 | Q1 | International Conference on Database Theory | 20 | 29 |
| 288 | Q1 | Euromicro International Conference on Parallel, Distributed and Network-Based Processing | 20 | 28 |
| 288 | Q1 | IEEE International Conference on Distributed Computing in Sensor Systems | 20 | 28 |
| 288 | Q1 | IEEE Symposium on Computers and Communications (ISCC) | 20 | 28 |
| 288 | Q1 | International Conference on Affective Computing and Intelligent Interaction and Workshops | 20 | 28 |
| 292 | Q1 | International Joint Conference on Natural Language Processing (IJCNLP) | 20 | 27 |
| 293 | Q1 | IEEE Radar Conference | 20 | 26 |
| 293 | Q1 | International Conference on Electrical Machines (ICEM) | 20 | 26 |
| 293 | Q1 | International Conference on Parallel problem solving from nature | 20 | 26 |
| 296 | Q1 | IAPR International Workshop on Document Analysis Systems | 20 | 25 |
| 296 | Q1 | IEEE Military Communications Conference | 20 | 25 |
| 296 | Q1 | International Symposium on Modeling and Optimization in Mobile, Ad Hoc, and Wireless Networks and Workshops | 20 | 25 |
| 296 | Q1 | Mathematical Foundations of Computer Science (MFCS) | 20 | 25 |
| 300 | Q1 | Nordic Conference on Human-Computer Interaction: Fun, Fast, Foundational | 20 | 24 |
| 301 | Q1 | IEEE Workshop on Automatic Speech Recognition & Understanding | 19 | 34 |
| 302 | Q1 | IEEE Vehicular Networking Conference (VNC) | 19 | 32 |
| 302 | Q1 | International Symposium on Symbolic and Algebraic Computation | 19 | 32 |
| 304 | Q1 | Annual ACM workshop on Privacy in the electronic society | 19 | 29 |
| 304 | Q1 | Artificial Intelligence and Interactive Digital Entertainment Conference | 19 | 29 |
| 304 | Q1 | International Conference on Recent Trends in Information, Telecommunication and Computing | 19 | 29 |
| 307 | Q1 | IFIP Conference on Human-Computer Interaction (INTERACT) | 19 | 28 |
| 307 | Q1 | International Conference on Web Engineering (ICWE) | 19 | 28 |
| 309 | Q1 | IEEE International Conference on Rehabilitation Robotics | 19 | 27 |
| 309 | Q1 | IEEE International Symposium on Workload Characterization | 19 | 27 |
| 311 | Q1 | International Conference on Intelligent Tutoring Systems | 19 | 26 |
| 311 | Q1 | International Symposium on Distributed Computing (DISC) | 19 | 26 |
| 311 | Q1 | Symposium on VLSI Circuits (VLSIC) | 19 | 26 |
| 314 | Q1 | Conference on Local Computer Networks (LCN) | 19 | 25 |
| 314 | Q1 | European Conference on Software Architecture | 19 | 25 |
| 314 | Q1 | IEEE Education Engineering (EDUCON) | 19 | 25 |
| 314 | Q1 | International Conference on Conceptual modeling | 19 | 25 |
| 314 | Q1 | International Conference on Database Systems for Advanced Applications | 19 | 25 |
| 319 | Q1 | European Signal Processing Conference (EUSIPCO) | 19 | 24 |
| 319 | Q1 | IEEE Antennas and Propagation Society International Symposium | 19 | 24 |
| 319 | Q1 | International Conference on Availability, Reliability and Security, ARES | 19 | 24 |
| 319 | Q1 | International Conference on Foundations of Software Science and Computation Structures (FoSSaCS) | 19 | 24 |
| 319 | Q1 | Logic Programming and Automated Reasoning (RCLP/LPAR) | 19 | 24 |
| 319 | Q1 | Principles and Practice of Constraint Programming (CP) | 19 | 24 |
| 325 | Q1 | ACM International Health Informatics Symposium | 19 | 23 |
| 326 | Q1 | International Conference on Software Testing, Verification, and Validation Workshops | 18 | 35 |
| 327 | Q1 | IEEE/WIC/ACM International Conference on Web Intelligence and Intelligent Agent Technology | 18 | 33 |
| 328 | Q1 | ACM Workshop on Embedded Sensing Systems for Energy-Efficiency in Buildings | 18 | 29 |
| 328 | Q1 | Annual ACM Web Science Conference | 18 | 29 |
| 328 | Q1 | International Symposium on Wireless Communication Systems | 18 | 29 |
| 328 | Q1 | Knowledge Engineering and Management by the Masses (EKAW) | 18 | 29 |
| 332 | Q1 | IEEE Spoken Language Technology Workshop (SLT) | 18 | 28 |
| 332 | Q1 | Power Electronics and Motion Control Conference | 18 | 28 |
| 334 | Q1 | Annual Meeting of the Special Interest Group on Discourse and Dialogue (SIGDIAL) | 18 | 27 |
| 335 | Q1 | IEEE Conference on Industrial Electronics and Applications | 18 | 26 |
| 335 | Q1 | IEEE International Conference on Trust, Security and Privacy in Computing and Communications (TrustCom) | 18 | 26 |
| 335 | Q1 | IFIP/IEEE International Symposium on Integrated Network Management | 18 | 26 |
| 335 | Q1 | International Conference on Electronics Computer Technology (ICECT) | 18 | 26 |
| 339 | Q1 | IEEE/ACM International Conference on Grid Computing | 18 | 25 |
| 339 | Q1 | Integer Programming and Combinatorial Optimization (IPCO) | 18 | 25 |
| 339 | Q1 | International Symposium on Personal, Indoor and Mobile Radio Communications Workshops (PIMRC Workshops) | 18 | 25 |
| 339 | Q1 | International Workshop on Wearable and Implantable Body Sensor Networks | 18 | 25 |
| 343 | Q1 | Approximation Algorithms for Combinatorial Optimization (APPROX) | 18 | 24 |
| 343 | Q1 | IEEE International Conference on Wireless and Mobile Computing Networking and Communications | 18 | 24 |
| 343 | Q1 | IEEE/RAS-EMBS International Conference on Biomedical Robotics and Biomechatronics | 18 | 24 |
| 343 | Q1 | International Conference on Parallel and Distributed Systems | 18 | 24 |
| 343 | Q1 | Workshop on Statistical Machine Translation | 18 | 24 |
| 348 | Q1 | Automated Technology for Verification and Analysis | 18 | 23 |
| 348 | Q1 | IEEE International Symposium on Power Line Communications and Its Applications | 18 | 23 |
| 348 | Q1 | International Workshop on Internet and Network Economics | 18 | 23 |
| 348 | Q1 | Symposium on Graph Drawing (GD) | 18 | 23 |
| 352 | Q1 | International Conference on Complex, Intelligent and Software Intensive Systems | 18 | 22 |
| 352 | Q1 | International Conference on Wireless On-Demand Network Systems and Services | 18 | 22 |
| 354 | Q1 | IEEE International Conference on Development and Learning (ICDL) | 18 | 21 |
| 354 | Q1 | International Symposium on Quality Electronic Design | 18 | 21 |
| 356 | Q1 | IEEE International Conference on Semantic Computing | 18 | 20 |
| 356 | Q1 | International IEEE Enterprise Distributed Object Computing Conference, EDOC | 18 | 20 |
| 358 | Q1 | Design Science Research in Information Systems and Technologies (DESRIST) | 17 | 37 |
| 359 | Q1 | Statistical and Scientific Database Management (SSDBM) | 17 | 31 |
| 360 | Q1 | IEEE/ACM International Conference on Utility and Cloud Computing | 17 | 30 |
| 361 | Q1 | International Conference on Transparent Optical Networks | 17 | 28 |
| 362 | Q1 | ACM International Conference on Modeling, Analysis and Simulation of Wireless and Mobile Systems | 17 | 27 |
| 362 | Q1 | Symposium on Haptic Interfaces for Virtual Environment and Teleoperator Systems | 17 | 27 |
| 364 | Q1 | IEEE Workshop on Signal Processing Advances in Wireless Communications | 17 | 26 |
| 364 | Q1 | International Conference on Information processing in medical imaging | 17 | 26 |
| 366 | Q1 | ACM International Conference on Computing Frontiers | 17 | 24 |
| 366 | Q1 | Foundations of Software Technology and Theoretical Computer Science (FSTTCS) | 17 | 24 |
| 366 | Q1 | International Conference on Communication Systems and Network Technologies (CSNT) | 17 | 24 |
| 366 | Q1 | International Conference on Frontiers in Handwriting Recognition | 17 | 24 |
| 366 | Q1 | International Symposium on Experimental Robotics | 17 | 24 |
| 366 | Q1 | Symposium on Reliable Distributed Systems (SRDS) | 17 | 24 |
| 366 | Q1 | Working Conference on Advanced Visual Interfaces (AVI) | 17 | 24 |
| 373 | Q2 | ACM SIGCHI Symposium on Engineering interactive computing systems | 17 | 23 |
| 373 | Q2 | IEEE International Electric Machines & Drives Conference | 17 | 23 |
| 373 | Q2 | IEEE International Symposium on Power Semiconductor Devices and ICs | 17 | 23 |
| 373 | Q2 | IEEE Workshop on Multimedia Signal Processing | 17 | 23 |
| 373 | Q2 | International Conference on Cryptology in Africa | 17 | 23 |
| 373 | Q2 | International Symposium on Power Electronics, Electrical Drives, Automation and Motion | 17 | 23 |
| 373 | Q2 | International Symposium on Wireless Pervasive Computing | 17 | 23 |
| 373 | Q2 | VLSI Test Symposium (VTS) | 17 | 23 |
| 381 | Q2 | IEEE Control Applications,(CCA) & Intelligent Control,(ISIC) | 17 | 22 |
| 381 | Q2 | IEEE International Midwest Symposium on Circuits and Systems | 17 | 22 |
| 381 | Q2 | IEEE International Symposium on Power Electronics for Distributed Generation Systems | 17 | 22 |
| 381 | Q2 | International Conference on Computational Linguistics and Intelligent Text Processing | 17 | 22 |
| 381 | Q2 | International Conference on Requirements Engineering: Foundation for Software Quality | 17 | 22 |
| 381 | Q2 | International Conference on VLSI Design | 17 | 22 |
| 381 | Q2 | International Conference on Wireless Communications, Networking and Mobile Computing | 17 | 22 |
| 381 | Q2 | International Symposium on Communications, Control and Signal Processing | 17 | 22 |
| 389 | Q2 | European Conference on Antennas and Propagation | 17 | 21 |
| 389 | Q2 | IEEE International Conference on Computer Science and Information Technology (ICCSIT) | 17 | 21 |
| 389 | Q2 | IEEE/AIAA Digital Avionics Systems Conference (DASC) | 17 | 21 |
| 389 | Q2 | International Conference on Distributed Computing Systems Workshops, ICDCSW | 17 | 21 |
| 389 | Q2 | International Workshop on Quality of Service (IWQoS) | 17 | 21 |
| 394 | Q2 | Biennial IEEE Conference on Electromagnetic Field Computation | 17 | 20 |
| 394 | Q2 | The Florida AI Research Society (FLAIRS) | 17 | 20 |
| 394 | Q2 | Web3D / VRML Symposium | 17 | 20 |
| 397 | Q2 | IEEE Asian Solid-State Circuits Conference, A-SSCC | 17 | 19 |
| 398 | Q2 | IEEE International Workshop on Information Forensics and Security | 16 | 31 |
| 399 | Q2 | International Symposium on Physical Design | 16 | 29 |
| 399 | Q2 | International Workshop on Quality of Multimedia Experience | 16 | 29 |
| 401 | Q2 | IEEE International Conference on Digital Ecosystems and Technologies, DEST | 16 | 28 |
| 402 | Q2 | IEEE International Conference on Communication Software and Networks (ICCSN) | 16 | 27 |
| 402 | Q2 | International ITG Workshop on Smart Antennas (WSA) | 16 | 27 |
| 404 | Q2 | International Conference on Wireless Communications and Signal Processing | 16 | 26 |
| 405 | Q2 | Annual Conference on Computer Graphics (SIGGRAPH) | 16 | 25 |
| 406 | Q2 | IEEE International Conference on Software Engineering and Formal Methods | 16 | 24 |
| 406 | Q2 | IEEE International Conference on Telecommunications (ICT) | 16 | 24 |
| 406 | Q2 | International Conference on Machine Learning and Cybernetics | 16 | 24 |
| 409 | Q2 | DAGM Symposium for Pattern Recognition | 16 | 23 |
| 409 | Q2 | Data Compression Conference, DCC | 16 | 23 |
| 409 | Q2 | European Conference on Synthetic Aperture Radar | 16 | 23 |
| 409 | Q2 | European Microwave Conference | 16 | 23 |
| 409 | Q2 | IEEE International Conference on Computer Design, ICCD | 16 | 23 |
| 409 | Q2 | IEEE International Symposium on Wearable Computers | 16 | 23 |
| 409 | Q2 | IEEE Intersociety Conference on Thermal and Thermomechanical Phenomena in Electronic Systems (ITherm) | 16 | 23 |
| 409 | Q2 | International Conference on Natural Computation | 16 | 23 |
| 409 | Q2 | Pacific Asia Conference on Information Systems (PACIS) | 16 | 23 |
| 418 | Q2 | IEEE International Conference on Industrial Informatics | 16 | 22 |
| 418 | Q2 | International Conference on Artificial Intelligence and Soft Computing ICAISC | 16 | 22 |
| 418 | Q2 | International Conference on Computational Problem-Solving (ICCP) | 16 | 22 |
| 418 | Q2 | International Conference on Computer Safety, Reliability, and Security | 16 | 22 |
| 418 | Q2 | International Conference on Digital Signal Processing | 16 | 22 |
| 418 | Q2 | Rewriting Techniques and Applications (RTA) | 16 | 22 |
| 424 | Q2 | IEEE International Conference on Advanced Learning Technologies | 16 | 21 |
| 424 | Q2 | IEEE International Conference on Tools with Artificial Intelligence | 16 | 21 |
| 424 | Q2 | International Conference on Computing, Networking and Communications | 16 | 21 |
| 427 | Q2 | IEEE International Conference on Robotics and Biomimetics | 16 | 20 |
| 427 | Q2 | IEEE International Electric Vehicle Conference (IEVC) | 16 | 20 |
| 427 | Q2 | IEEE Radio and Wireless Symposium | 16 | 20 |
| 427 | Q2 | IEEE/ASME International Conference on Advanced Intelligent Mechatronics | 16 | 20 |
| 427 | Q2 | IEEE/IFIP International Conference on Dependable Systems and Networks Workshops (DSN-W) | 16 | 20 |
| 427 | Q2 | International Conference on High Performance Computing | 16 | 20 |
| 427 | Q2 | International Conference on Innovative Mobile and Internet Services in Ubiquitous Computing | 16 | 20 |
| 427 | Q2 | International Conference on Machine Learning and Applications | 16 | 20 |
| 427 | Q2 | Joint Conference on Innovation and Technology in Computer Science Education | 16 | 20 |
| 436 | Q2 | IEEE Symposium on 3D User Interfaces | 16 | 19 |
| 436 | Q2 | IET International Conference on Power Electronics, Machines and Drives | 16 | 19 |
| 436 | Q2 | International Conference on Computer and Communication Technology (ICCCT) | 16 | 19 |
| 439 | Q2 | Euromicro Conference on Digital System Design, Architectures, Methods and Tools, DSD | 16 | 18 |
| 439 | Q2 | International Conference on Electrical Machines and Systems | 16 | 18 |
| 439 | Q2 | International Conference on Embedded Computer Systems (SAMOS) | 16 | 18 |
| 442 | Q2 | European Conference on Towards a service-based internet | 15 | 28 |
| 442 | Q2 | International Conference on Network-Based Information Systems | 15 | 28 |
| 444 | Q2 | IEEE Workshop on Applications of Signal Processing to Audio and Acoustics | 15 | 27 |
| 444 | Q2 | International Conference on Cloud Computing and Services Science (CLOSER) | 15 | 27 |
| 444 | Q2 | International Solid-State Sensors, Actuators and Microsystems Conference (TRANSDUCERS) | 15 | 27 |
| 444 | Q2 | Latin American Theoretical INformatics (LATIN) | 15 | 27 |
| 448 | Q2 | International Conference on Intelligent Computation Technology and Automation | 15 | 25 |
| 449 | Q2 | Conference on Web Accessibility | 15 | 24 |
| 450 | Q2 | IEEE Computer Society Annual Symposium on VLSI | 15 | 23 |
| 450 | Q2 | IEEE Symposium on Visual Languages and Human-Centric Computing | 15 | 23 |
| 450 | Q2 | International Conference on Fuzzy Systems and Knowledge Discovery | 15 | 23 |
| 453 | Q2 | Artificial Intelligence and Symbolic Computation (AISC) | 15 | 22 |
| 453 | Q2 | European Solid State Device Research Conference, ESSDERC | 15 | 22 |
| 453 | Q2 | IEEE International Conference on Engineering of Complex Computer Systems | 15 | 22 |
| 453 | Q2 | IEEE Singapore International Conference on Communication Systems | 15 | 22 |
| 453 | Q2 | Integration of AI and OR Techniques in Contraint Programming for Combinatorial Optimzation Problem (CPAIOR) | 15 | 22 |
| 453 | Q2 | International Conference on Intelligent Information Hiding and Multimedia Signal Processing | 15 | 22 |
| 453 | Q2 | International Workshop on Parameterized and Exact Computation | 15 | 22 |
| 453 | Q2 | Workshop on Positioning, Navigation and Communication | 15 | 22 |
| 461 | Q2 | International Conference on Advanced Computer Theory and Engineering (ICACTE) | 15 | 21 |
| 461 | Q2 | International Conference on Computer Application and System Modeling (ICCASM) | 15 | 21 |
| 461 | Q2 | International Conference on Control, Automation, Robotics and Vision | 15 | 21 |
| 461 | Q2 | International Conference on Cryptology in India (INDOCRYPT) | 15 | 21 |
| 461 | Q2 | International Conference on Environment and Electrical Engineering (EEEIC) | 15 | 21 |
| 461 | Q2 | International Conference on Network and System Security | 15 | 21 |
| 461 | Q2 | International Conference on Signal Processing | 15 | 21 |
| 461 | Q2 | International Conference on Swarm Intelligence | 15 | 21 |
| 461 | Q2 | International Symposium on Algorithms and Computation (ISAAC) | 15 | 21 |
| 461 | Q2 | International Symposium on Parallel and Distributed Processing with Applications | 15 | 21 |
| 461 | Q2 | International Symposium on Robotics - ISR | 15 | 21 |
| 472 | Q2 | ACM Great Lakes Symposium on VLSI | 15 | 20 |
| 472 | Q2 | Conference on Multimedia Modeling | 15 | 20 |
| 472 | Q2 | Frontiers in Education Conference | 15 | 20 |
| 472 | Q2 | IEEE International Workshop on Safety, Security and Rescue Robotics | 15 | 20 |
| 472 | Q2 | International Conference on Current Trends in Theory and Practice of Computer Science | 15 | 20 |
| 472 | Q2 | International Conference on Intelligent Sensors, Sensor Networks and Information Processing | 15 | 20 |
| 472 | Q2 | International Congress on Image and Signal Processing, CISP | 15 | 20 |
| 472 | Q2 | International Workshop on Cognitive Information Processing | 15 | 20 |
| 472 | Q2 | WICSA/ECSA Joint Working IEEE/IFIP Conference on Software Architecture | 15 | 20 |
| 481 | Q2 | Algorithmic Learning Theory (ALT) | 15 | 19 |
| 481 | Q2 | European Symposium on Artificial Neural Networks (ESANN) | 15 | 19 |
| 481 | Q2 | IEEE Compound Semiconductor Integrated Circuit Symposium (CSICS) | 15 | 19 |
| 481 | Q2 | IEEE Electrical Power & Energy Conference (EPEC) | 15 | 19 |
| 481 | Q2 | International Conference of Distributed Computing and Networking | 15 | 19 |
| 481 | Q2 | International Conference on Database and expert systems applications | 15 | 19 |
| 481 | Q2 | International Conference on Optimization of Electrical and Electronic Equipment | 15 | 19 |
| 481 | Q2 | International Symposium on Visual Computing | 15 | 19 |
| 489 | Q2 | Chinese Control and Decision Conference | 15 | 18 |
| 489 | Q2 | IEEE International Conference on Embedded and Real-Time Computing Systems and Applications (RTCSA) | 15 | 18 |
| 489 | Q2 | International Conference on Cognitive Radio Oriented Wireless Networks and Communications | 15 | 18 |
| 489 | Q2 | International Conference on Emerging Trends in Electrical and Computer Technology (ICETECT) | 15 | 18 |
| 489 | Q2 | International Conference on Intelligent Systems Design and Applications | 15 | 18 |
| 489 | Q2 | International Conference on Mechatronics and Automation | 15 | 18 |
| 489 | Q2 | WRI International Conference on Communications and Mobile Computing | 15 | 18 |
| 496 | Q2 | IEEE International Symposium on Object Oriented Real-Time Distributed Computing | 15 | 17 |
| 496 | Q2 | IEEE International Workshop on Machine Learning for Signal Processing | 15 | 17 |
| 496 | Q2 | International Conference on Quality Software | 15 | 17 |
| 499 | Q2 | International Conference on Artificial Neural Networks | 14 | 30 |
| 500 | Q2 | IEEE International Symposium on Consumer Electronics | 14 | 24 |
| 500 | Q2 | International Workshop on Education Technology and Computer Science (ETCS) | 14 | 24 |
| 502 | Q2 | ACM Symposium on Computing for Development | 14 | 23 |
| 502 | Q2 | Australasian Conference on Information Security and Privacy (ACISP) | 14 | 23 |
| 502 | Q2 | European Conference on Interactive tv and video | 14 | 23 |
| 502 | Q2 | International Conference on the Synthesis and Simulation of Living Systems | 14 | 23 |
| 506 | Q2 | EUROMICRO Conference on Software Engineering and Advanced Applications | 14 | 22 |
| 506 | Q2 | International Conference on Compilers, Architecture, and Synthesis for Embedded Systems (CASES) | 14 | 22 |
| 506 | Q2 | International Conference on Intelligent Systems, Modelling and Simulation | 14 | 22 |
| 509 | Q2 | IEEE Asia-Pacific Services Computing Conference, APSCC | 14 | 21 |
| 509 | Q2 | IEEE International Conference on Bioinformatics and Biomedicine | 14 | 21 |
| 509 | Q2 | IEEE International Conference on Global Software Engineering | 14 | 21 |
| 509 | Q2 | IEEE Statistical Signal Processing Workshop (SSP) | 14 | 21 |
| 509 | Q2 | International Conference on Hybrid Artificial Intelligence Systems | 14 | 21 |
| 509 | Q2 | International Conference on Image Analysis and Recognition | 14 | 21 |
| 509 | Q2 | International Symposium on Communication Systems Networks and Digital Signal Processing | 14 | 21 |
| 516 | Q2 | Data Warehousing and Knowledge Discovery (DaWaK) | 14 | 20 |
| 516 | Q2 | European Conference on Evolutionary Computation in Combinatorial Optimization | 14 | 20 |
| 516 | Q2 | European Conference on Logics in Artificial Intelligence | 14 | 20 |
| 516 | Q2 | IEEE Wireless Communications and Networking Conference Workshops (WCNCW) | 14 | 20 |
| 516 | Q2 | IFIP International Information Security and Privacy Conference | 14 | 20 |
| 516 | Q2 | International Conference on Computer Communication and Informatics | 14 | 20 |
| 516 | Q2 | International Conference on Electrical Engineering/Electronics, Computer, Telecommunications and Information Technology | 14 | 20 |
| 516 | Q2 | International Conference on Electronics and Information Engineering (ICEIE) | 14 | 20 |
| 524 | Q2 | IEEE International Conference on Cloud Computing and Intelligence Systems | 14 | 19 |
| 524 | Q2 | International Conference on Artificial Intelligence and Statistics | 14 | 19 |
| 524 | Q2 | International Conference on Industrial engineering and other applications of applied intelligent systems | 14 | 19 |
| 524 | Q2 | International Conference on Measuring Technology and Mechatronics Automation | 14 | 19 |
| 524 | Q2 | International Conference on Ubiquitous and Future Networks | 14 | 19 |
| 524 | Q2 | Workshop in Information Security Theory and Practice (WISTP) | 14 | 19 |
| 530 | Q2 | Conference on Control and Fault-Tolerant Systems (SysTol) | 14 | 18 |
| 530 | Q2 | European Conference on Mobile Robots | 14 | 18 |
| 530 | Q2 | IEEE Semiconductor Thermal Measurement and Management Symposium | 14 | 18 |
| 530 | Q2 | IEEE/ACM International Symposium on Nanoscale Architectures | 14 | 18 |
| 530 | Q2 | International Conference on Advanced Computer Control (ICACC) | 14 | 18 |
| 530 | Q2 | International Conference on Automotive User Interfaces and Interactive Vehicular Applications | 14 | 18 |
| 530 | Q2 | International Conference on Broadband, Wireless Computing, Communication and Applications | 14 | 18 |
| 530 | Q2 | International Conference on Computational Intelligence and Communication Networks | 14 | 18 |
| 530 | Q2 | International Conference on Knowledge-based and intelligent information and engineering systems | 14 | 18 |
| 530 | Q2 | International Conference on Neural Information Processing | 14 | 18 |
| 530 | Q2 | International Conference On Principles Of DIstributed Systems | 14 | 18 |
| 540 | Q2 | ACM Symposium on Document Engineering | 14 | 17 |
| 540 | Q2 | Asian Conference on Intelligent Information and Database Systems | 14 | 17 |
| 540 | Q2 | Australian Computer-Human Interaction Conference | 14 | 17 |
| 540 | Q2 | IEEE International Advance Computing Conference | 14 | 17 |
| 540 | Q2 | IEEE International Conference on Intelligence and Security Informatics | 14 | 17 |
| 540 | Q2 | IEEE International Symposium on Design and Diagnostics of Electronic Circuits & Systems | 14 | 17 |
| 540 | Q2 | International ACM Sigsoft Symposium on Component-based software engineering | 14 | 17 |
| 540 | Q2 | International Conference on Computers Helping People with Special Needs | 14 | 17 |
| 540 | Q2 | International Conference on Education Technology and Computer | 14 | 17 |
| 540 | Q2 | International Conference on Electrical and Control Engineering (ICECE) | 14 | 17 |
| 540 | Q2 | International Conference on Haptics-generating and perceiving tangible sensations | 14 | 17 |
| 540 | Q2 | Web Information Systems Engineering (WISE) | 14 | 17 |
| 552 | Q2 | International Conference on Intelligent Networking and Collaborative Systems (INCoS) | 14 | 16 |
| 552 | Q2 | International Conference on ITS Telecommunications | 14 | 16 |
| 552 | Q2 | International Conference on Ultra Modern Telecommunications & Workshops ICUMT | 14 | 16 |
| 552 | Q2 | International Multi-Conference on Systems, Signals and Devices (SSD) | 14 | 16 |
| 552 | Q2 | Workshop on Graph-Theoretic Concepts in Computer Science (WG) | 14 | 16 |
| 557 | Q2 | IEEE International Symposium on Multimedia | 14 | 15 |
| 557 | Q2 | IEEE/IFIP International Conference on VLSI and System-on-Chip (VLSI-SoC) | 14 | 15 |
| 557 | Q2 | International Conference on Information Visualization | 14 | 15 |
| 560 | Q2 | International Conference on Information Security | 13 | 26 |
| 561 | Q2 | International Conference on Advanced Computing & Communication Technologies | 13 | 25 |
| 562 | Q2 | International Conference on Communications and Networking in China | 13 | 24 |
| 562 | Q2 | Rough Sets and Current Trends in Computing (RSCTC) | 13 | 24 |
| 564 | Q2 | Conference on Privacy, Security and Trust | 13 | 23 |
| 564 | Q2 | Electronics Packaging Technology Conference | 13 | 23 |
| 564 | Q2 | IEEE International Working Conference on Source Code Analysis and Manipulation | 13 | 23 |
| 564 | Q2 | International Workshop on Spoken Language Translation | 13 | 23 |
| 568 | Q2 | ACM International Conference on Supporting group work | 13 | 21 |
| 568 | Q2 | International Conference on New Technologies, Mobility and Security | 13 | 21 |
| 570 | Q2 | Asian Conference on Programming languages and systems | 13 | 20 |
| 570 | Q2 | International Conference on Algorithms and Architectures for Parallel Processing | 13 | 20 |
| 570 | Q2 | International Conference on Computational Intelligence, Communication Systems and Networks | 13 | 20 |
| 570 | Q2 | Logic Programming and Non-Monotonic Reasoning (LPNMR) | 13 | 20 |
| 570 | Q2 | SIBGRAPI Conference on Graphics, Patterns and Images | 13 | 20 |
| 575 | Q2 | European Conference on Genetic Programming | 13 | 19 |
| 575 | Q2 | IEEE Visual Communications and Image Processing (VCIP) | 13 | 19 |
| 575 | Q2 | IEEE Workshop on Control and Modeling for Power Electronics (COMPEL) | 13 | 19 |
| 575 | Q2 | IEEE/ACM International Conference on Formal Methods and Models for Codesign | 13 | 19 |
| 575 | Q2 | Innovative Applications of Artificial Intelligence (IAAI) | 13 | 19 |
| 575 | Q2 | International Conference on Computer Science and Education (ICCSE) | 13 | 19 |
| 575 | Q2 | International Conference on Networks Security, Wireless Communications and Trusted Computing | 13 | 19 |
| 582 | Q2 | ACM International Conference on Bioinformatics and Computational Biology | 13 | 18 |
| 582 | Q2 | Applications of Natural Language to Data Bases (NLDB) | 13 | 18 |
| 582 | Q2 | Asia-Pacific Software Engineering Conference | 13 | 18 |
| 582 | Q2 | Colloquium on Structural Information & Communication Complexity | 13 | 18 |
| 582 | Q2 | Conference on Wireless Health | 13 | 18 |
| 582 | Q2 | European Radar Conference | 13 | 18 |
| 582 | Q2 | IEEE International Conference on e-Health Networking Applications and Services | 13 | 18 |
| 582 | Q2 | IEEE International Conference on Mechatronics | 13 | 18 |
| 582 | Q2 | IEEE International Conference on Networking, Sensing and Control | 13 | 18 |
| 582 | Q2 | IEEE International Conference on Software Engineering and Service Sciences | 13 | 18 |
| 582 | Q2 | IEEE International Conference on Wireless, Mobile and Ubiquitous Technologies in Education | 13 | 18 |
| 582 | Q2 | IEEE International Symposium on Computer-Based Medical Systems, CBMS | 13 | 18 |
| 582 | Q2 | International Conference on Computational Science (ICCS) | 13 | 18 |
| 582 | Q2 | International Conference on Computing and Combinatorics | 13 | 18 |
| 582 | Q2 | International Conference on Evaluation and Assessment in Software Engineering | 13 | 18 |
| 582 | Q2 | International Symposium on Applied Sciences in Biomedical and Communication Technologies (ISABEL) | 13 | 18 |
| 582 | Q2 | International Workshop on Computing Education Research | 13 | 18 |
| 582 | Q2 | Scandinavian Workshop on Algorithm Theory (SWAT) | 13 | 18 |
| 600 | Q2 | Annual Mediterranean Ad Hoc Networking Workshop (Med-Hoc-Net) | 13 | 17 |
| 600 | Q2 | Antennas & Propagation Conference | 13 | 17 |
| 600 | Q2 | IEEE Biomedical Circuits and Systems Conference (BioCAS) | 13 | 17 |
| 600 | Q2 | IEEE Conference on Robotics, Automation and Mechatronics | 13 | 17 |
| 600 | Q2 | IEEE International Conference on Communication Technology (ICCT) | 13 | 17 |
| 600 | Q2 | International Conference on Computer and Network Technology | 13 | 17 |
| 600 | Q2 | International Conference on Industrial and Information Systems (IIS) | 13 | 17 |
| 600 | Q2 | International Conference on Information Assurance and Security | 13 | 17 |
| 600 | Q2 | International Symposium on Turbo Codes and Iterative Information Processing | 13 | 17 |
| 609 | Q2 | IEEE International Conference and Workshops on the Engineering of Computer-Based Systems | 13 | 16 |
| 609 | Q2 | IEEE International Workshop on Computational Advances in Multi-Sensor Adaptive Processing | 13 | 16 |
| 609 | Q2 | IEEE Workshop on Signal Processing Systems | 13 | 16 |
| 609 | Q2 | IEEE/IFIP International Conference on Embedded and Ubiquitous Computing | 13 | 16 |
| 609 | Q2 | Information Security and Cryptology (ICISC) | 13 | 16 |
| 609 | Q2 | International Conference for Internet Technology and Secured Transactions | 13 | 16 |
| 609 | Q2 | International Conference on Agents and Artificial Intelligence (ICAART) | 13 | 16 |
| 609 | Q2 | International Conference on Computing Communication and Networking Technologies (ICCCNT) | 13 | 16 |
| 609 | Q2 | International Conference on Control, Automation and Systems, ICCAS | 13 | 16 |
| 609 | Q2 | International Conference on Multimedia Information Networking and Security | 13 | 16 |
| 609 | Q2 | International Conference on Software, Telecommunications and Computer Networks | 13 | 16 |
| 609 | Q2 | International Conference on the Quality of Information and Communications Technology | 13 | 16 |
| 609 | Q2 | International Symposium on Wireless Personal Multimedia Communications | 13 | 16 |
| 622 | Q2 | Device Research Conference | 13 | 15 |
| 623 | Q2 | IEEE International Symposium on Broadband Multimedia Systems and Broadcasting | 13 | 14 |
| 623 | Q2 | International Conference on Computer Vision and Graphics | 13 | 14 |
| 623 | Q2 | Iranian Conference on Electrical Engineering | 13 | 14 |
| 626 | Q2 | International Conference on Computer and Communication Engineering | 12 | 32 |
| 627 | Q2 | International Telecommunications Energy Conference | 12 | 25 |
| 628 | Q2 | Annual International Conference on Digital Government Research | 12 | 24 |
| 629 | Q2 | IEEE International Symposium on Network Computing and Applications | 12 | 22 |
| 630 | Q2 | Asia Communications and Photonics Conference (ACP) | 12 | 21 |
| 630 | Q2 | IEEE International Conference on Wireless Information Technology and Systems | 12 | 21 |
| 632 | Q2 | Annual IFIP WG 11.3 Conference on Data and applications security and privacy | 12 | 20 |
| 632 | Q2 | International Colloquium on Signal Processing & Its Applications | 12 | 20 |
| 632 | Q2 | Joint International Conference on Power Electronics, Drives and Energy Systems (PEDES) | 12 | 20 |
| 633 | Q2 | Electronic Government and the Information Systems Perspective (EGOVIS) | 12 | 19 |
| 633 | Q2 | International Conference on Systems, Signals and Image Processing | 12 | 19 |
| 633 | Q2 | International Conference on Wireless Communication, Vehicular Technology, Information Theory and Aerospace & Electronic Systems Technology Wireless VITAE | 12 | 19 |
| 636 | Q2 | IEEE International Conference on Electronics, Circuits and Systems | 12 | 18 |
| 636 | Q2 | IEEE International Conference on Multisensor Fusion and Integration for Intelligent Systems | 12 | 18 |
| 636 | Q2 | IEEE International Conference on Secure Software Integration and Reliability Improvement SSIRI | 12 | 18 |
| 636 | Q2 | International Conference on Advances in computer science and information technology | 12 | 18 |
| 636 | Q2 | International Conference on Clean Electrical Power | 12 | 18 |
| 636 | Q2 | International Conference on Computer Supported Cooperative Work in Design, CSCWD | 12 | 18 |
| 636 | Q2 | International Conference on Consumer Electronics, Communications and Networks | 12 | 18 |
| 636 | Q2 | International Conference on Intelligence in Next Generation Networks (ICIN) | 12 | 18 |
| 636 | Q2 | Wireless Telecommunications Symposium, WTS | 12 | 18 |
| 645 | Q2 | ACM Symposium on Virtual Reality Software and Technology | 12 | 17 |
| 645 | Q2 | ACM/IEEE International Conference on Distributed Smart Cameras | 12 | 17 |
| 645 | Q2 | IEEE International Conference on Granular Computing | 12 | 17 |
| 645 | Q2 | IEEE International Power Electronics and Motion Control Conference | 12 | 17 |
| 645 | Q2 | International Conference on Cloud and Service Computing | 12 | 17 |
| 645 | Q2 | International Conference on Computer analysis of images and patterns | 12 | 17 |
| 645 | Q2 | International Conference on Computing for Geospatial Research & Applications | 12 | 17 |
| 645 | Q2 | International Conference on Enterprise Information Systems (ICEIS) | 12 | 17 |
| 645 | Q2 | International Conference on Search based software engineering | 12 | 17 |
| 645 | Q2 | International Symposium on Applications and the Internet | 12 | 17 |
| 645 | Q2 | International Topical Meeting on & Microwave Photonics Conference, Asia-Pacific, MWP/APMP | 12 | 17 |
| 645 | Q2 | International Workshop on Database and Expert Systems Applications | 12 | 17 |
| 645 | Q2 | Scandinavian Conference on Image Analysis | 12 | 17 |
| 658 | Q2 | Advances in Databases and Information Systems (ADBIS) | 12 | 16 |
| 658 | Q2 | Canadian Conference on Computer and Robot Vision (CRV) | 12 | 16 |
| 658 | Q2 | European Microwave Integrated Circuits Conference | 12 | 16 |
| 658 | Q2 | Federated Conference on Computer Science and Information Systems | 12 | 16 |
| 658 | Q2 | IEEE International Conference on Wireless Communications, Networking and Information Security | 12 | 16 |
| 658 | Q2 | IEEE International Symposium on Computer-Aided Control System Design (CACSD) | 12 | 16 |
| 658 | Q2 | International Conference on Bioinformatics and Biomedical Engineering | 12 | 16 |
| 658 | Q2 | International Conference on Cyber-Enabled Distributed Computing and Knowledge Discovery (CyberC) | 12 | 16 |
| 658 | Q2 | International Conference on Digital Image Computing Techniques and Applications (DICTA) | 12 | 16 |
| 658 | Q2 | International Conference on Web-age information management | 12 | 16 |
| 658 | Q2 | International Congress on Mathematical Software (ICMS) | 12 | 16 |
| 658 | Q2 | International Radar Symposium | 12 | 16 |
| 658 | Q2 | International Symposium on Autonomous Decentralized Systems | 12 | 16 |
| 671 | Q2 | 3DTV Conference: The True Vision-Capture, Transmission and Display of 3D Video (3DTV-CON) | 12 | 15 |
| 674 | Q2 | ACIS International Conference on Software Engineering, Artificial Intelligence, Networking, and Parallel/Distributed Computing | 12 | 15 |
| 674 | Q2 | Australasian Conference on Artificial Intelligence | 12 | 15 |
| 674 | Q2 | IEEE Conference on Technologies for Homeland Security | 12 | 15 |
| 674 | Q2 | IEEE International Performance, Computing, and Communications Conference, IPCCC | 12 | 15 |
| 674 | Q2 | IEEE MTT-S International Microwave Workshop Series on Innovative Wireless Power Transmission: Technologies, Systems, and Applications (IMWS) | 12 | 15 |
| 674 | Q2 | IEEE/ACS International Conference on Computer Systems and Applications | 12 | 15 |
| 674 | Q2 | IEEE-CS Conference on Software Engineering Education and Training (CSEE&T) | 12 | 15 |
| 674 | Q2 | International Conference on Emerging Trends in Engineering and Technology | 12 | 15 |
| 674 | Q2 | International Conference on Image Information Processing (ICIIP) | 12 | 15 |
| 674 | Q2 | International Conference on Information Networking (ICOIN) | 12 | 15 |
| 674 | Q2 | International Conference on Intelligent Control and Information Processing (ICICIP) | 12 | 15 |
| 674 | Q2 | International Conference on Scale Space and Variational Methods in Computer Vision | 12 | 15 |
| 674 | Q2 | International Conference on Software Technology and Engineering | 12 | 15 |
| 674 | Q2 | International Workshop on Content-Based Multimedia Indexing | 12 | 15 |
| 674 | Q2 | International Workshop on Intelligent Systems and Applications | 12 | 15 |
| 674 | Q2 | Simulation of Adaptive Behavior | 12 | 15 |
| 674 | Q2 | Text, Speech and Dialogue (TSD) | 12 | 15 |
| 674 | Q2 | Workshop on Hyperspectral Image and Signal Processing: Evolution in Remote Sensing (WHISPERS) | 12 | 15 |
| 692 | Q2 | Asia-Pacific Microwave Conference | 12 | 14 |
| 692 | Q2 | Iberian Conference on Information Systems and Technologies | 12 | 14 |
| 692 | Q2 | India Software Engineering Conference | 12 | 14 |
| 692 | Q2 | International Conference on Devices and Communications (ICDeCom) | 12 | 14 |
| 692 | Q2 | International Conference on Image Analysis and Signal Processing | 12 | 14 |
| 692 | Q2 | Pacific Rim International Symposium on Dependable Computing | 12 | 14 |
| 698 | Q2 | Mexican International Conference on Artificial Intelligence (MICAI) | 12 | 13 |
| 699 | Q2 | IEEE Symposium on Computational Intelligence and Data Mining | 11 | 23 |
| 700 | Q2 | IEEE International Conference on Cognitive Informatics (ICCI) | 11 | 21 |
| 700 | Q2 | International Conference on Networking and Computing | 11 | 21 |
| 700 | Q2 | Wireless and Optical Communications Conference (WOCC) | 11 | 21 |
| 703 | Q2 | IEEE International Conference on Computational Intelligence and Computing Research (ICCIC) | 11 | 20 |
| 703 | Q2 | Joint IFIP Wireless and Mobile Networking Conference (WMNC) | 11 | 20 |
| 705 | Q2 | IEEE Convention of Electrical and Electronics Engineers in Israel (IEEEI) | 11 | 19 |
| 705 | Q2 | International Conference on Signal Processing Systems | 11 | 19 |
| 705 | Q2 | International Symposium on Computer Architecture and High Performance Computing, SBAC- | 11 | 19 |
| | | PAD | | |
| 705 | Q2 | International Workshop on Cellular Neural Networks and Their Applications | 11 | 19 |
| 709 | Q2 | Asia-Pacific Web Conference | 11 | 18 |
| 709 | Q2 | Canadian Conference on Advances in Artificial Intelligence | 11 | 18 |
| 709 | Q2 | Education in the Knowledge Society (EKS) | 11 | 18 |
| 709 | Q2 | German Conference on Advances in artificial intelligence | 11 | 18 |
| 709 | Q2 | IEEE International Workshop on Computer Aided Modeling and Design of Communication Links and Networks (CAMAD) | 11 | 18 |
| 709 | Q2 | International Conference on Electrical Engineering and Informatics (ICEEI) | 11 | 18 |
| 709 | Q2 | International Conference on Green Circuits and Systems | 11 | 18 |
| 709 | Q2 | International Conference on Information and Communications Security | 11 | 18 |
| 709 | Q2 | International Conference on Networking, Architecture, and Storage | 11 | 18 |
| 709 | Q2 | International Conference on Swarm, Evolutionary, and Memetic Computing | 11 | 18 |
| 719 | Q2 | Biennial Symposium on Communications (QBSC) | 11 | 17 |
| 719 | Q2 | IEEE International Vacuum Electronics Conference (IVEC) | 11 | 17 |
| 719 | Q2 | IEEE Symposium on Industrial Electronics & Applications | 11 | 17 |
| 719 | Q2 | International Conference on Body Area Networks | 11 | 17 |
| 719 | Q2 | International Conference on Circuit and Signal Processing | 11 | 17 |
| 719 | Q2 | International Conference on Communication, Computing & Security | 11 | 17 |
| 719 | Q2 | International Conference on Computer Information Systems and Industrial Management Applications | 11 | 17 |
| 719 | Q2 | International Conference on High performance computing for computational science | 11 | 17 |
| 719 | Q2 | International Conference on Information Integration and Web-based Applications & Services (IIWAS) | 11 | 17 |
| 719 | Q2 | International Conference on Intelligent Human-Machine Systems and Cybernetics | 11 | 17 |
| 719 | Q2 | International Workshop on Antenna Technology: Small Antennas and Novel Metamaterials | 11 | 17 |
| 730 | Q2 | IEEE Bipolar/BiCMOS Circuits and Technology Meeting (BCTM) | 11 | 16 |
| 730 | Q2 | IEEE International Conference on Electro/Information Technology | 11 | 16 |
| 730 | Q2 | IEEE International Conference on Intelligent Computing and Intelligent Systems | 11 | 16 |
| 730 | Q2 | IEEE International Symposium on Precision Clock Synchronization for Measurement, Control and Communication | 11 | 16 |
| 730 | Q2 | IEEE/ACIS International Conference on Computer and Information Science ICIS | 11 | 16 |
| 730 | Q2 | Industrial Conference on Data Mining | 11 | 16 |
| 730 | Q2 | International Conference on Advances in Computer Entertainment | 11 | 16 |
| 730 | Q2 | International Conference on Biomedical Engineering and Informatics | 11 | 16 |
| 730 | Q2 | International Conference on Computational Intelligence and Security (CIS) | 11 | 16 |
| 730 | Q2 | International Conference on Logic Programming | 11 | 16 |
| 730 | Q2 | International Symposium on Resilient Control Systems | 11 | 16 |
| 730 | Q2 | International Symposium on VLSI Design, Automation and Test | 11 | 16 |
| 730 | Q2 | International Workshop on Knowledge Discovery and Data Mining | 11 | 16 |
| 730 | Q2 | Proceedings of the International Conference on Bio-inspired Systems and Signal Processing (BIOSIGNALS) | 11 | 16 |
| 730 | Q2 | Product Focused Software Process Improvement (PROFES) | 11 | 16 |
| 730 | Q2 | Workshop on Algorithms and Data Structures (WADS) | 11 | 16 |
| 746 | Q3 | ACM International Conference on Underwater Networks and Systems | 11 | 15 |
| 746 | Q3 | IFIP International Conference on Wireless and Optical Communications Networks | 11 | 15 |
| 746 | Q3 | International Conference on Advanced Robotics | 11 | 15 |
| 746 | Q3 | International Conference on Computational Aspects of Social Networks | 11 | 15 |
| 746 | Q3 | International Conference on Computational Intelligence and Software Engineering, CiSE | 11 | 15 |
| 746 | Q3 | International Conference on Computer Science and Network Technology (ICCSNT) | 11 | 15 |
| 746 | Q3 | International Conference on Intelligent Engineering Systems | 11 | 15 |
| 746 | Q3 | International Conference on Next Generation Mobile Applications, Services and Technologies | 11 | 15 |
| 746 | Q3 | International Symposium on Distributed Computing and Applications to Business, Engineering and Science | 11 | 15 |
| 746 | Q3 | International Symposium on Industrial Embedded Systems | 11 | 15 |
| 746 | Q3 | International Symposium on Intelligent Systems and Informatics | 11 | 15 |
| 746 | Q3 | International Workshop on Image Analysis for Multimedia Interactive Services | 11 | 15 |
| 746 | Q3 | NASA/ESA Conference on Adaptive Hardware and Systems, AHS | 11 | 15 |
| 746 | Q3 | Power Electronics, Drive Systems and Technologies Conference (PEDSTC) | 11 | 15 |
| 746 | Q3 | Workshop on Algorithms and Computation | 11 | 15 |
| 761 | Q3 | Electronic Commerce and Web Technologies (EC-Web) | 11 | 14 |
| 761 | Q3 | Electronics System-Integration Technology Conference | 11 | 14 |
| 761 | Q3 | European Conference on Power Electronics and Applications | 11 | 14 |
| 761 | Q3 | IEEE International Conference on Big Data | 11 | 14 |
| 761 | Q3 | IEEE International Conference on Bioinformatics and Biomedicine Workshops (BIBMW) | 11 | 14 |
| 761 | Q3 | International Conference on Applied Robotics for the Power Industry | 11 | 14 |
| 761 | Q3 | International Conference on Artificial Intelligence and Computational Intelligence, AICI | 11 | 14 |
| 761 | Q3 | International Conference on Combinatorial algorithms | 11 | 14 |
| 761 | Q3 | International Conference on Electric Utility Deregulation and Restructuring and Power Technologies (DRPT) | 11 | 14 |
| 761 | Q3 | International Conference on Image Processing Theory Tools and Applications (IPTA) | 11 | 14 |
| 761 | Q3 | International Conference on Microelectronics | 11 | 14 |
| 761 | Q3 | International Conference on Networking and Information Technology | 11 | 14 |
| 761 | Q3 | International Conference on Software Business (ICSOB) | 11 | 14 |
| 761 | Q3 | International Conference on Web-Based Learning | 11 | 14 |
| 761 | Q3 | International Symposium on Communications and Information Technologies | 11 | 14 |
| 761 | Q3 | International Symposium on Innovations in Intelligent Systems and Applications (INISTA) | 11 | 14 |
| 761 | Q3 | World Congress on Information and Communication Technologies (WICT) | 11 | 14 |
| 761 | Q3 | World Congress on Nature & Biologically Inspired Computing, NaBIC | 11 | 14 |
| 779 | Q3 | AAAI Conference on Human Computation and Crowdsourcing | 11 | 13 |
| 779 | Q3 | Advances in Mobile Multimedia (MoMM) | 11 | 13 |
| 779 | Q3 | Canadian Conference on Computational Geometry | 11 | 13 |
| 779 | Q3 | IEEE Asia Pacific Conference on Circuits and Systems, APCCAS | 11 | 13 |
| 779 | Q3 | IEEE Global Conference on Signal and Information Processing (GlobalSIP) | 11 | 13 |
| 779 | Q3 | IEEE International Conference on Bio-Inspired Computing: Theories and Applications | 11 | 13 |
| 779 | Q3 | IEEE International Conference on Broadband Network and Multimedia Technology (IC-BNMT) | 11 | 13 |
| 779 | Q3 | IEEE International Symposium on Applied Computational Intelligence and Informatics (SACI) | 11 | 13 |
| 779 | Q3 | IEEE International Symposium on Defect and Fault Tolerance of VLSI Systems, DFTVS | 11 | 13 |
| 779 | Q3 | IEEE International Symposium on Diagnostics for Electric Machines, Power Electronics & Drives (SDEMPED) | 11 | 13 |
| 779 | Q3 | IEEE International Symposium on Signal Processing and Information Technology | 11 | 13 |
| 779 | Q3 | International Conference on Electric Information and Control Engineering (ICEICE) | 11 | 13 |
| 779 | Q3 | International Conference on Future Computer and Communication (ICFCC) | 11 | 13 |
| 779 | Q3 | International Conference on Information Technology and Applications in Biomedicine | 11 | 13 |
| 779 | Q3 | International Symposium on Applied Machine Intelligence and Informatics | 11 | 13 |
| 794 | Q3 | IEEE Holm Conference on Electrical Contacts | 11 | 12 |
| 794 | Q3 | IEEE Workshop on Local & Metropolitan Area Networks | 11 | 12 |
| 794 | Q3 | International Conference on Genetic and Evolutionary Computing (ICGEC) | 11 | 12 |
| 794 | Q3 | International Conference on Multimedia Technology | 11 | 12 |
| 794 | Q3 | International Conference on Thermal, Mechanical and Multiphysics Simulation and Experiments in Micro-Electronics and Micro-Systems | 11 | 12 |
| 799 | Q3 | International Symposium on Symbolic and Numeric Algorithms for Scientific Computing | 11 | 11 |
| 800 | Q3 | Information Security for South Africa (ISSA) | 10 | 22 |
| 801 | Q3 | International Conference on Computer Science and Electronics Engineering | 10 | 21 |
| 801 | Q3 | Saudi International Electronics, Communications and Photonics Conference (SIECPC) | 10 | 21 |
| 803 | Q3 | International Conference on Computational Intelligence, Modelling and Simulation (CIMSiM) | 10 | 20 |
| 804 | Q3 | IEEE International Conferences on Internet of Things, and Cyber, Physical and Social Computing | 10 | 19 |
| 804 | Q3 | International Conference on Computational Processing of the Portuguese Language | 10 | 19 |
| 804 | Q3 | International Conference on Hardware and Software: verification and testing | 10 | 19 |
| 807 | Q3 | IFIP TC 6/TC International Conference on Communications and Multimedia Security | 10 | 18 |
| 807 | Q3 | International Conference on Signal Processing and Communications | 10 | 18 |
| 807 | Q3 | International Symposium on Temporal Representation and Reasoning | 10 | 18 |
| 810 | Q3 | European Workshop on Visual Information Processing | 10 | 17 |
| 810 | Q3 | Hellenic Conference on Artificial Intelligence (SETN) | 10 | 17 |
| 810 | Q3 | IEEE Topical Meeting on Silicon Monolithic Integrated Circuits in RF Systems (SiRF) | 10 | 17 |
| 810 | Q3 | International Conference on Electronics, Communications and Control (ICECC) | 10 | 17 |
| 810 | Q3 | International Conference on Interactive Digital Storytelling | 10 | 17 |
| 810 | Q3 | International Conference on Software and System Process | 10 | 17 |
| 816 | Q3 | Conference on Information Security and Cryptology (INSCRYPT) | 10 | 16 |
| 816 | Q3 | IEEE International Symposium on Approximate Dynamic Programming and Reinforcement Learning | 10 | 16 |
| 816 | Q3 | IEEE International Symposium on Electronic Design, Test and Applications | 10 | 16 |
| 816 | Q3 | IEEE Topical Conference on Biomedical Wireless Technologies, Networks, and Sensing Systems (BioWireleSS) | 10 | 16 |
| 816 | Q3 | International Conference on Electronic and Mechanical Engineering and Information Technology (EMEIT) | 10 | 16 |
| 816 | Q3 | International Conference on Information security applications | 10 | 16 |
| 816 | Q3 | International Conference on Information Systems Security | 10 | 16 |
| 816 | Q3 | International Conference on Solid-State and Integrated Circuit Technology | 10 | 16 |
| 816 | Q3 | International Conference on Web Information Systems and Mining | 10 | 16 |
| 816 | Q3 | International Workshop on Reconfigurable Communication-centric Systems-on-Chip | 10 | 16 |
| 826 | Q3 | European Conference on Software Process Improvement | 10 | 15 |
| 826 | Q3 | Iberoamerican Congress on Pattern Recognition CIARP | 10 | 15 |
| 826 | Q3 | IEEE International Conference on Computer Science and Automation Engineering (CSAE) | 10 | 15 |
| 826 | Q3 | IEEE/SEMI Advanced Semiconductor Manufacturing Conference, ASMC | 10 | 15 |
| 826 | Q3 | International Conference on Advances in Recent Technologies in Communication and Computing | 10 | 15 |
| 826 | Q3 | International Conference on Digital Information Management | 10 | 15 |
| 826 | Q3 | International Conference on Educational and Information Technology (ICEIT) | 10 | 15 |
| 826 | Q3 | International Conference on Parallel and Distributed Computing, Applications and Technologies | 10 | 15 |
| 826 | Q3 | International Conference on Semantics, Knowledge and Grid | 10 | 15 |
| 826 | Q3 | International Symposium on Electronic System Design | 10 | 15 |
| 836 | Q3 | ACIS International Conference on Software Engineering Research, Management & Applications | 10 | 14 |
| 836 | Q3 | Canadian Conference on Computer Science & Software Engineering | 10 | 14 |
| 836 | Q3 | IEEE Annual Wireless and Microwave Technology Conference | 10 | 14 |
| 836 | Q3 | IEEE Electric Ship Technologies Symposium | 10 | 14 |
| 836 | Q3 | IEEE International Conference on Intelligent Data Acquisition and Advanced Computing Systems (IDAACS) | 10 | 14 |
| 836 | Q3 | IEEE International Conference on Network Infrastructure and Digital Content | 10 | 14 |
| 836 | Q3 | IEEE International Conference on Vehicular Electronics and Safety | 10 | 14 |
| 836 | Q3 | IEEE Symposium on Computers & Informatics (ISCI) | 10 | 14 |
| 836 | Q3 | International Asia Conference on Informatics in Control, Automation and Robotics | 10 | 14 |
| 836 | Q3 | International Conference on Computer Systems and Technologies | 10 | 14 |
| 836 | Q3 | International Conference on Discovery science | 10 | 14 |
| 836 | Q3 | International Conference on Emerging Security Technologies | 10 | 14 |
| 836 | Q3 | International Conference on Image and Signal Processing | 10 | 14 |
| 836 | Q3 | International Conference on Intelligent Computing and Integrated Systems (ICISS) | 10 | 14 |
| 836 | Q3 | International Conference on Mobile Business | 10 | 14 |
| 836 | Q3 | International Workshop on Structural and Syntactic Pattern Recognition (SSPR) | 10 | 14 |
| 836 | Q3 | Workshop on Approximation and Online Algorithms | 10 | 14 |
| 853 | Q3 | Asia-Pacific Signal and Information Processing Association Annual Summit and Conference (APSIPA) | 10 | 13 |
| 853 | Q3 | Biennial Baltic Electronics Conference (BEC) | 10 | 13 |
| 853 | Q3 | IEEE International Conference on Control System, Computing and Engineering (ICCSCE) | 10 | 13 |
| 853 | Q3 | IEEE International Symposium on Electrical Insulation, ISEI | 10 | 13 |
| 853 | Q3 | International Conference on Advances in Intelligent Data Analysis | 10 | 13 |
| 853 | Q3 | International Conference on Computer, Mechatronics, Control and Electronic Engineering | 10 | 13 |
| 853 | Q3 | International Conference on Hybrid Intelligent Systems | 10 | 13 |
| 853 | Q3 | International Conference on Image and Graphics | 10 | 13 |
| 853 | Q3 | International Conference on Informatics in Control, Automation and Robotics | 10 | 13 |
| 853 | Q3 | International Conference on Internet Computing & Information Services (ICICIS) | 10 | 13 |
| 853 | Q3 | International Conference on Multimedia Computing and Systems | 10 | 13 |
| 853 | Q3 | International Conference on Parallel Processing and Applied Mathematics | 10 | 13 |
| 853 | Q3 | International Conference on Real-Time Networks and Systems | 10 | 13 |
| 853 | Q3 | International Conference on Theory and Practice of Electronic Governance | 10 | 13 |
| 853 | Q3 | International Conference-Workshop on Compatibility and Power Electronics (CPE) | 10 | 13 |
| 853 | Q3 | International IEEE Conference on Signal-Image Technologies and Internet-Based System | 10 | 13 |
| 853 | Q3 | International Symposium on Neural Networks: Advances in Neural Networks | 10 | 13 |
| 853 | Q3 | Meeting of the North American Fuzzy Information Processing Society, NAFIPS | 10 | 13 |
| 871 | Q3 | ACM International Conference on Design of Communication | 10 | 12 |
| 871 | Q3 | IEEE Conference on Electrical Performance of Electronic Packaging and Systems (EPEPS) | 10 | 12 |
| 871 | Q3 | IEEE International New Circuits and Systems Conference | 10 | 12 |
| 871 | Q3 | International Conference on Adaptive and natural computing algorithms | 10 | 12 |
| 871 | Q3 | International Conference on Computer Supported Education - CSEDU | 10 | 12 |
| 871 | Q3 | International Conference on Computing, Electronics and Electrical Technologies | 10 | 12 |
| 871 | Q3 | International Conference on Fun with Algorithms | 10 | 12 |
| 871 | Q3 | International Conference on Intelligent Robotics and Applications | 10 | 12 |
| 871 | Q3 | International Conference on Knowledge Engineering and Ontology Development (KEOD) | 10 | 12 |
| 871 | Q3 | International Conference on Legal Knowledge and Information Systems (JURIX) | 10 | 12 |
| 871 | Q3 | International Conference on Mechanic Automation and Control Engineering | 10 | 12 |
| 871 | Q3 | International Conference on Methods and Models in Automation and Robotics | 10 | 12 |
| 871 | Q3 | International Conference on Pattern Recognition, Informatics and Mobile Engineering | 10 | 12 |
| 871 | Q3 | International Conference on Power Engineering, Energy and Electrical Drives | 10 | 12 |
| 871 | Q3 | International Conference on Software Engineering and Data Mining | 10 | 12 |
| 871 | Q3 | International Symposium on Computer, Communication, Control and Automation | 10 | 12 |
| 871 | Q3 | International Symposium on Image and Signal Processing and Analysis, ISPA | 10 | 12 |
| 871 | Q3 | International Symposium on Intelligent Information Technology and Security Informatics | 10 | 12 |
| 871 | Q3 | International Symposium on Telecommunications | 10 | 12 |
| 871 | Q3 | International Syposium on Methodologies for Intelligent Systems (ISMIS) | 10 | 12 |
| 871 | Q3 | International Workshop on Advanced Computational Intelligence (IWACI) | 10 | 12 |
| 871 | Q3 | International Workshop on Inductive Logic Programming (ILP) | 10 | 12 |
| 871 | Q3 | Joint Urban Remote Sensing Event (JURSE) | 10 | 12 |
| 871 | Q3 | Symposium on Integrated Circuits and Systems Design | 10 | 12 |
| 895 | Q3 | IEEE International Conference on Bioinformatics and Bioengineering, BIBE | 10 | 11 |
| 895 | Q3 | IEEE International Conference on Computing, Control and Industrial Engineering (CCIE) | 10 | 11 |
| 895 | Q3 | International Conference on Advances in Computing, Communications and Informatics | 10 | 11 |
| 895 | Q3 | International Conference on Electronic Packaging Technology & High Density Packaging | 10 | 11 |
| 895 | Q3 | International Conference on Information Retrieval & Knowledge Management | 10 | 11 |
| 895 | Q3 | International Conference on Management of Emergent Digital EcoSystems | 10 | 11 |
| 895 | Q3 | International Conference on MultiMedia and Information Technology | 10 | 11 |
| 895 | Q3 | International Conference on Power, Control and Embedded Systems | 10 | 11 |
| 895 | Q3 | International Work-Conference on Artificial and Natural Neural Networks (IWANN) | 10 | 11 |
| 895 | Q3 | Wireless Advanced (WiAd) | 10 | 11 |
| 895 | Q3 | Workshop on Logic, Language, Information and Computation (WoLLIC) | 10 | 11 |
| 906 | Q3 | Anais do Simpósio Brasileiro de Informática na Educação | 10 | 10 |
| 906 | Q3 | International Conference on Microwave and Millimeter Wave Technology | 10 | 10 |
| 906 | Q3 | MESA IEEE/ASME International Conference on Mechtronic and Embedded Systems and Applications | 10 | 10 |
| 909 | Q3 | Annual Conference on Information technology education | 9 | 21 |
| 910 | Q3 | International Conference on Internet Technology and Applications (iTAP) | 9 | 20 |
| 911 | Q3 | International Conference on Advances in information and computer security | 9 | 19 |
| 912 | Q3 | Workshop on Cyber Security and Information Intelligence Research | 9 | 19 |
| 913 | Q3 | IEEE International Conference on Software Security and Reliability Companion | 9 | 18 |
| 913 | Q3 | International Conference on Knowledge Management and Knowledge Technologies | 9 | 18 |
| 915 | Q3 | International Colloquium on Theoretical Aspects of Computing | 9 | 17 |
| 915 | Q3 | International Conference on Mechanical and Electrical Technology | 9 | 17 |
| 915 | Q3 | WSEAS International Conference on Automatic control, modelling & simulation | 9 | 17 |
| 918 | Q3 | CIRP-Sponsored International Conference on Digital Enterprise Technology | 9 | 16 |
| 918 | Q3 | International Conference on Information, Communications & Signal Processing | 9 | 16 |
| 918 | Q3 | International Conference on Signal Processing and Communication Systems | 9 | 16 |
| 918 | Q3 | International Symposium on Medical Information and Communication Technology | 9 | 16 |
| 918 | Q3 | Pacific Rim Conference on Multimedia | 9 | 16 |
| 923 | Q3 | Digital Information and Communication Technology and Its Applications (DICTAP) | 9 | 15 |
| 923 | Q3 | European Intelligence and Security Informatics Conference | 9 | 15 |
| 923 | Q3 | IEEE International Conference on Digital Game and Intelligent Toy Enhanced Learning | 9 | 15 |
| 923 | Q3 | IEEE Pacific Rim Conference on Communications, Computers and Signal Processing PacRim | 9 | 15 |
| 923 | Q3 | IFIP International Conference on Network and Parallel Computing | 9 | 15 |
| 923 | Q3 | International Conference on Mobile Ad-hoc and Sensor Networks | 9 | 15 |
| 923 | Q3 | Pacific Rim International Conference on Artificial Intelligence (PRICAI) | 9 | 15 |
| 930 | Q3 | IEEE International Conference on Grey Systems and Intelligent Services | 9 | 14 |
| 930 | Q3 | Indian Conference on Computer Vision, Graphics & Image Processing, ICVGIP | 9 | 14 |
| 930 | Q3 | Intelligent Data Engineering and Automated Learning (IDEAL) | 9 | 14 |
| 930 | Q3 | International Conference on Autonomous Robots and Agents | 9 | 14 |
| 930 | Q3 | International Conference on Communications and Signal Processing (ICCSP) | 9 | 14 |
| 930 | Q3 | International Conference on Health Informatics (HEALTHINF) | 9 | 14 |
| 930 | Q3 | International Conference on Image Analysis and Processing, ICIAP | 9 | 14 |
| 930 | Q3 | International Conference on Natural Language Processing and Knowledge Engineering | 9 | 14 |
| 930 | Q3 | International Symposium on Environmental Software Systems (ISESS) | 9 | 14 |
| 930 | Q3 | International Symposium on the Physical and Failure Analysis of Integrated Circuits | 9 | 14 |
| 930 | Q3 | International Workshop on Systems, Signal Processing and their Applications | 9 | 14 |
| 930 | Q3 | Mexican Conference on Pattern Recognition | 9 | 14 |
| 930 | Q3 | The European Conference on Antennas and Propagation: EuCAP | 9 | 14 |
| 943 | Q3 | Australian Communications Theory Workshop (AusCTW) | 9 | 13 |
| 943 | Q3 | IEEE Conference on Information & Communication Technologies | 9 | 13 |
| 943 | Q3 | IEEE International Carnahan Conference on Security Technology, ICCST | 9 | 13 |
| 943 | Q3 | IEEE International Conference on Automation, Quality and Testing, Robotics, AQTR | 9 | 13 |
| 943 | Q3 | IEEE International Conference on Information Theory and Information Security | 9 | 13 |
| 943 | Q3 | IEEE International Conference on Service Operations and Logistics, and Informatics | 9 | 13 |
| 943 | Q3 | IEEE International Conference on Signal Processing, Communications and Computing | 9 | 13 |
| 943 | Q3 | International Conference on Communications, Circuits and Systems, ICCCAS | 9 | 13 |
| 943 | Q3 | International Conference on Computer Applications and Industrial Electronics (ICCAIE) | 9 | 13 |
| 943 | Q3 | International Conference on Computer Science and Service System (CSSS) | 9 | 13 |
| 943 | Q3 | International Conference on Computers & Industrial Engineering | 9 | 13 |
| 943 | Q3 | International Conference on E-Health Networking, Digital Ecosystems and Technologies (EDT) | 9 | 13 |
| 943 | Q3 | International Conference on Electrical and Computer Engineering (ICECE) | 9 | 13 |
| 943 | Q3 | International Conference on Electrical Power Quality and Utilisation | 9 | 13 |
| 943 | Q3 | International Conference on Process Automation, Control and Computing | 9 | 13 |
| 943 | Q3 | International Conference on Signal Processing, Communication, Computing and Networking Technologies | 9 | 13 |
| 943 | Q3 | International Conference on Software and Data Technologies | 9 | 13 |
| 943 | Q3 | International Conference on Virtual Systems and Multimedia | 9 | 13 |
| 943 | Q3 | International Database Engineering & Applications Symposium | 9 | 13 |
| 943 | Q3 | International IEEE Conference on Intelligent Systems | 9 | 13 |
| 943 | Q3 | International Symposium on Information Theory and Its Applications | 9 | 13 |
| 943 | Q3 | International Symposium on Intelligent Signal Processing and Communication Systems | 9 | 13 |
| 943 | Q3 | International Symposium on VLSI Technology, Systems, and Applications | 9 | 13 |
| 943 | Q3 | International Wireless Internet Conference | 9 | 13 |
| 943 | Q3 | Koli Calling International Conference on Computing Education Research | 9 | 13 |
| 943 | Q3 | OptoElectronics and Communications Conference OECC | 9 | 13 |
| 943 | Q3 | Workshop on Advanced Issues of E-Commerce and Web/based Information Systems (WECWIS) | 9 | 13 |
| 943 | Q3 | WSEAS International Conference on Artificial intelligence, knowledge engineering and data bases | 9 | 13 |
| 971 | Q3 | ACM International Workshop on Mobility Management and Wireless Access, MOBIWAC | 9 | 12 |
| 971 | Q3 | IAPR Asian Conference on Pattern Recognition | 9 | 12 |
| 971 | Q3 | Iberian Conference on Pattern Recognition and Image Analysis | 9 | 12 |
| 971 | Q3 | IEEE International Conference on Computational Cybernetics | 9 | 12 |
| 971 | Q3 | IEEE International Conference on Computing and Communication Technologies, Research, Innovation, and Vision for the Future (RIVF) | 9 | 12 |
| 971 | Q3 | IEEE International Conference on Intelligent Computer Communication and Processing | 9 | 12 |
| 971 | Q3 | IEEE International Conference on Progress in Informatics and Computing | 9 | 12 |
| 971 | Q3 | IEEE International Conference on Technologies for Practical Robot Applications | 9 | 12 |
| 971 | Q3 | IEEE International Symposium on Computational Intelligence and Informatics (CINTI) | 9 | 12 |
| 971 | Q3 | IEEE International Workshop on Haptic Audio visual Environments and Games | 9 | 12 |
| 971 | Q3 | International Conference Image and Vision Computing New Zealand | 9 | 12 |
| 971 | Q3 | International Conference of Soft Computing and Pattern Recognition | 9 | 12 |
| 971 | Q3 | International Conference on Cloud and Green Computing | 9 | 12 |
| 971 | Q3 | International Conference on Computer Games | 9 | 12 |
| 971 | Q3 | International Conference on Electronic Devices, Systems and Applications (ICEDSA) | 9 | 12 |
| 971 | Q3 | International Conference on Ground Penetrating Radar | 9 | 12 |
| 971 | Q3 | International Conference on Multimedia and Ubiquitous Engineering | 9 | 12 |
| 971 | Q3 | International Conference on Wireless Mobile Communication and Healthcare (Mobihealth) | 9 | 12 |
| 971 | Q3 | International Symposium on Performance Evaluation of Computer and Telecommunication Systems | 9 | 12 |
| 971 | Q3 | Telecommunications Forum (TELFOR) | 9 | 12 |
| 991 | Q3 | Asia Communications and Photonics Conference and Exhibition | 9 | 11 |
| 991 | Q3 | Asia Information Retrieval Symposium | 9 | 11 |
| 991 | Q3 | Business Informatics Research (BIR) | 9 | 11 |
| 991 | Q3 | Conference on Colour in Graphics, Imaging, and Vision | 9 | 11 |
| 991 | Q3 | Conference on Local Computer Networks Workshops (LCN Workshops) | 9 | 11 |
| 991 | Q3 | Discrete Geometry for Computer Imagery (DGCI) | 9 | 11 |
| 991 | Q3 | European Conference on Circuit Theory and Design, ECCTD | 9 | 11 |
| 991 | Q3 | European Conference on Networks and Optical Communications | 9 | 11 |
| 991 | Q3 | European Society for Fuzzy Logic and Technology (EUSFLAT) | 9 | 11 |
| 991 | Q3 | IEEE Applied Imagery Pattern Recognition Workshop, AIPR | 9 | 11 |
| 991 | Q3 | IEEE International Congress on Big Data | 9 | 11 |
| 991 | Q3 | IEEE International RF and Microwave Conference, RFM | 9 | 11 |
| 991 | Q3 | IEEE National Aerospace and Electronics Conference (NAECON) | 9 | 11 |
| 991 | Q3 | IEEE Recent Advances in Intelligent Computational Systems (RAICS) | 9 | 11 |
| 991 | Q3 | IEEE Students' Conference on Electrical, Electronics and Computer Science | 9 | 11 |
| 991 | Q3 | IEEE Symposium on Computational Intelligence in Bioinformatics and Computational Biology | 9 | 11 |
| 991 | Q3 | IEEE Topical Conference on Wireless Sensors and Sensor Networks (WiSNet) | 9 | 11 |
| 991 | Q3 | IEEE-EMBS International Conference on Biomedical and Health Informatics | 9 | 11 |
| 991 | Q3 | International Conference on Advanced data mining and applications | 9 | 11 |
| 991 | Q3 | International Conference on Artificial Intelligence, Management Science and Electronic Commerce | 9 | 11 |
| 991 | Q3 | International Conference on Autonomic and Autonomous Systems, ICAS | 9 | 11 |
| 991 | Q3 | International Conference on Cloud Networking (CloudNet) | 9 | 11 |
| 991 | Q3 | International Conference on Power Electronics and Drive Systems | 9 | 11 |
| 991 | Q3 | International Conference on Robotics in Alpe-Adria-Danube Region | 9 | 11 |
| 991 | Q3 | International Conference System Theory, Control and Computing | 9 | 11 |
| 991 | Q3 | International Symposium on Parallel and Distributed Computing | 9 | 11 |
| 991 | Q3 | Soft Computing Models in Industrial and Environmental Applications (SOCO) | 9 | 11 |
| 991 | Q3 | Workshop Methoden und Beschreibungssprachen zur Modellierung und Verifikation von Schaltungen und Systemen (MBMV) | 9 | 11 |
| 1019 | Q3 | ESA Workshop on Satellite Navigation Technologies and European Workshop on GNSS Signals and Signal Processing (NAVITEC) | 9 | 10 |
| 1019 | Q3 | IEEE International Symposium on IT in Medicine & Education | 9 | 10 |
| 1019 | Q3 | International Conference on Communications, Computing and Control Applications | 9 | 10 |
| 1019 | Q3 | International Conference on Industrial Mechatronics and Automation (ICIMA) | 9 | 10 |
| 1019 | Q3 | International Symposium on Antenna Technology and Applied Electromagnetics & the American Electromagnetics Conference (ANTEM-AMEREM) | 9 | 10 |
| 1024 | Q3 | IEEE Control and System Graduate Research Colloquium (ICSGRC) | 9 | 9 |
| 1025 | Q3 | International Conference on Artificial Intelligence (IC-AI) | 8 | 22 |
| 1026 | Q3 | Conference on Natural Language Processing (KONVENS) | 8 | 18 |
| 1026 | Q3 | International conference on Information systems, Technology, and Management (ICISTM) | 8 | 18 |
| 1028 | Q3 | ACIS/JNU International Conference on Computers, Networks, Systems and Industrial Engineering | 8 | 15 |
| 1028 | Q3 | ACM Conference on Creativity and cognition | 8 | 15 |
| 1028 | Q3 | IEEE International Conference on Integrated Circuit Design and Technology and Tutorial, ICICDT | 8 | 15 |
| 1031 | Q3 | Brazilian Power Electronics Conference (COBEP) | 8 | 14 |
| 1031 | Q3 | International Conference on Anti-counterfeiting, Security, and Identification in Communication | 8 | 14 |
| 1031 | Q3 | International Conference on Applied Electronics (AE) | 8 | 14 |
| 1031 | Q3 | International Conference on Computational Advances in Bio and Medical Sciences | 8 | 14 |
| 1031 | Q3 | International Conference on Electrical Energy Systems | 8 | 14 |
| 1031 | Q3 | International Conference on Emerging Intelligent Data and Web Technologies | 8 | 14 |
| 1031 | Q3 | International Conference on Pattern Recognition Applications and Methods | 8 | 14 |
| 1031 | Q3 | Mobile Networks and Management (MONAMI) | 8 | 14 |
| 1039 | Q3 | Bioinspired Models of Network, Information, and Computing Systems (BIONETICS) | 8 | 13 |
| 1039 | Q3 | IEEE International Conference on Signal Processing, Computing and Control | 8 | 13 |
| 1039 | Q3 | IEEE Latin-American Conference on Communications | 8 | 13 |
| 1039 | Q3 | IEEE Radiation Effects Data Workshop | 8 | 13 |
| 1039 | Q3 | IFIP International Conference on Digital Forensics | 8 | 13 |
| 1039 | Q3 | International Conference on Mechanical and Electronics Engineering | 8 | 13 |
| 1039 | Q3 | International Conference on Pervasive Computing and Applications | 8 | 13 |
| 1039 | Q3 | International Conference on Wavelet Analysis and Pattern Recognition | 8 | 13 |
| 1039 | Q3 | Pacific Asia Conference on Language, Information, and Computation (PACLIC) | 8 | 13 |
| 1039 | Q3 | UK Workshop on Computational Intelligence | 8 | 13 |
| 1049 | Q3 | Edutainment - International Conference on E-learning and Games | 8 | 12 |
| 1049 | Q3 | European Conference on Computer Performance Engineering | 8 | 12 |
| 1049 | Q3 | German Conference on Pattern Recognition (GCPR) | 8 | 12 |
| 1049 | Q3 | IEEE Workshop on Advanced Robotics and its Social Impacts (ARSO) | 8 | 12 |
| 1049 | Q3 | International Conference on Communication, Information & Computing Technology | 8 | 12 |
| 1049 | Q3 | International Conference on Computer and Communication Technologies in Agriculture Engineering (CCTAE) | 8 | 12 |
| 1049 | Q3 | International Conference on Digital Content, Multimedia Technology and its Applications | 8 | 12 |
| 1049 | Q3 | International Conference on e-Education, e-Business, e-Management, and e-Learning | 8 | 12 |
| 1049 | Q3 | International Conference on Emerging Trends in Robotics and Communication Technologies | 8 | 12 |
| 1049 | Q3 | International Conference on Knowledge Discovery and Information Retrieval (KDIR) | 8 | 12 |
| 1049 | Q3 | International Conference on Networking and Distributed Computing | 8 | 12 |
| 1049 | Q3 | International Conference on Supply Chain Management and Information Systems | 8 | 12 |
| 1049 | Q3 | International Joint Conference on Computational Intelligence | 8 | 12 |
| 1049 | Q3 | Iranian Machine Vision and Image Processing (MVIP) | 8 | 12 |
| 1063 | Q3 | IEEE Conference on Electron Devices and Solid-State Circuits, EDSSC | 8 | 11 |
| 1063 | Q3 | IEEE International Conference on Communications in China | 8 | 11 |
| 1063 | Q3 | IEEE International Conference on Microelectronic Test Structures | 8 | 11 |
| 1063 | Q3 | IEEE International Semiconductor Laser Conference (ISLC) | 8 | 11 |
| 1063 | Q3 | IEEE Latin American Symposium on Circuits and Systems (LASCAS) | 8 | 11 |
| 1063 | Q3 | International Computer Symposium (ICS) | 8 | 11 |
| 1063 | Q3 | International Conference on Brain informatics | 8 | 11 |
| 1063 | Q3 | International Conference on Computer Science and Software Engineering | 8 | 11 |
| 1063 | Q3 | International Conference on Electrical, Control and Computer Engineering | 8 | 11 |
| 1063 | Q3 | International Conference on Electronic government and the information systems perspective | 8 | 11 |
| 1063 | Q3 | International Conference on Information Technology Based Higher Education and Training | 8 | 11 |
| 1063 | Q3 | International Conference on Innovations in Bio-inspired Computing and Applications | 8 | 11 |
| 1063 | Q3 | International Conference on Mechatronic Science, Electric Engineering and Computer | 8 | 11 |
| 1063 | Q3 | International Conference on Microelectronics | 8 | 11 |
| 1063 | Q3 | International Conference on Microwave Radar and Wireless Communications | 8 | 11 |
| 1063 | Q3 | International Conference on Networking and Digital Society | 8 | 11 |
| 1063 | Q3 | International Semiconductor Conference | 8 | 11 |
| 1063 | Q3 | International Symposium on Signals Systems and Electronics | 8 | 11 |
| 1063 | Q3 | International Symposium on Signals, Circuits and Systems | 8 | 11 |
| 1063 | Q3 | International Telecommunications Network Strategy and Planning Symposium | 8 | 11 |
| 1063 | Q3 | Mediterrannean Microwave Symposium (MMS) | 8 | 11 |
| 1063 | Q3 | OptoElectronics and Communications Conference and Photonics in Switching | 8 | 11 |
| 1063 | Q3 | Pacific-Asia Conference on Circuits, Communications and System (PACCS) | 8 | 11 |
| 1086 | Q3 | Asia Symposium on Quality Electronic Design | 8 | 10 |
| 1086 | Q3 | Balkan Conference in Informatics | 8 | 10 |
| 1086 | Q3 | Brazilian Symposium on Human Factors in Computing Systems | 8 | 10 |
| 1086 | Q3 | IEEE International Conference on Cloud Engineering | 8 | 10 |
| 1086 | Q3 | IEEE International Conference on Cyber Technology in Automation, Control, and Intelligent Systems | 8 | 10 |
| 1086 | Q3 | IEEE International Conference on Teaching, Assessment and Learning for Engineering | 8 | 10 |
| 1086 | Q3 | IEEE Region 8 International Conference on Computational Technologies in Electrical and Electronics Engineering (SIBIRCON) | 8 | 10 |
| 1086 | Q3 | IEEE Symposium on Wireless Technology and Applications | 8 | 10 |
| 1086 | Q3 | International Conference on Audio Language and Image Processing (ICALIP) | 8 | 10 |
| 1086 | Q3 | International Conference on Electronics and Optoelectronics (ICEOE) | 8 | 10 |
| 1086 | Q3 | International Conference on Informatics, Electronics & Vision | 8 | 10 |
| 1086 | Q3 | International Conference on Information Networking and Automation (ICINA) | 8 | 10 |
| 1086 | Q3 | International Conference on Intelligent Networks and Intelligent Systems | 8 | 10 |
| 1086 | Q3 | International Conference on Knowledge and Systems Engineering | 8 | 10 |
| 1086 | Q3 | International Conference on Knowledge Management and Information Sharing | 8 | 10 |
| 1086 | Q3 | International Conference on Machine Vision and Human-Machine Interface | 8 | 10 |
| 1086 | Q3 | International Conference on Pervasive Computing Signal Processing and Applications | 8 | 10 |
| 1086 | Q3 | International Conference on Remote Engineering and Virtual Instrumentation | 8 | 10 |
| 1086 | Q3 | International Conference on Signal and Image Processing | 8 | 10 |
| 1086 | Q3 | International Conference on Systems and Informatics | 8 | 10 |
| 1086 | Q3 | International Conference on Technologies and Applications of Artificial Intelligence | 8 | 10 |
| 1086 | Q3 | International Conference on Uncertainty Reasoning and Knowledge Engineering | 8 | 10 |
| 1086 | Q3 | International Power Electronics and Motion Control Conference | 8 | 10 |
| 1086 | Q3 | International Spring Seminar on Electronics Technology | 8 | 10 |
| 1086 | Q3 | International Symposium on Advanced Networks and Telecommunication Systems | 8 | 10 |
| 1086 | Q3 | International Symposium on Information Processing (ISIP) | 8 | 10 |
| 1086 | Q3 | International Symposium on Parallel Architectures, Algorithms and Programming | 8 | 10 |
| 1086 | Q3 | International Vacuum Electron Sources Conference and Nanocarbon (IVESC) | 8 | 10 |
| 1086 | Q3 | International Workshop on Digital Watermarking | 8 | 10 |
| 1086 | Q3 | International Workshop on Robotic and Sensors Environments | 8 | 10 |
| 1086 | Q3 | Symposium on Neural Network Applications in Electrical Engineering (NEUREL) | 8 | 10 |
| 1117 | Q4 | IEEE International Conference on Healthcare Informatics, Imaging and Systems Biology | 8 | 9 |
| 1117 | Q4 | IEEE Joint International Information Technology and Artificial Intelligence Conference (ITAIC) | 8 | 9 |
| 1117 | Q4 | IEEE Southwest Symposium on Image Analysis & Interpretation (SSIAI) | 8 | 9 |
| 1117 | Q4 | IEEE-APS Topical Conference on Antennas and Propagation in Wireless Communications | 8 | 9 |
| 1117 | Q4 | Integrated Communications, Navigation and Surveillance Conference | 8 | 9 |
| 1117 | Q4 | International Conference on Asian Digital Libraries | 8 | 9 |
| 1117 | Q4 | International Conference on Computer Engineering & Systems | 8 | 9 |
| 1117 | Q4 | International Conference on Digital Manufacturing & Automation | 8 | 9 |
| 1117 | Q4 | International Symposium on Antennas, Propagation and EM Theory | 8 | 9 |
| 1117 | Q4 | International Symposium on Electronics and Telecommunications | 8 | 9 |
| 1117 | Q4 | Special Interest Group on Computer Personnel Research Annual Conference (SIGCPR) | 8 | 9 |
| 1117 | Q4 | Spring Conference on Computer Graphics | 8 | 9 |
| 1117 | Q4 | Symposium on Information and Communication Technology | 8 | 9 |
| 1130 | Q4 | International Conference on Education and New Learning Technologies | 8 | 8 |
| 1130 | Q4 | International Conference on Ubiquitous Robots and Ambient Intelligence | 8 | 8 |
| 1130 | Q4 | International Electric Drives Production Conference (EDPC) | 8 | 8 |
| 1133 | Q4 | Evaluation of Novel Approaches to Software Engineering (ENASE) | 7 | 25 |
| 1134 | Q4 | International Conference on Intelligent System and Knowledge Engineering | 7 | 16 |
| 1134 | Q4 | Machine Learning and Data Mining in Pattern Recognition (MLDM) | 7 | 16 |
| 1136 | Q4 | Pacific-Rim Symposium on Image and Video Technology | 7 | 15 |
| 1137 | Q4 | International Conference on Bioinformatics and Biomedical Technology (ICBBT) | 7 | 14 |
| 1137 | Q4 | International Conference on Challenges in Environmental Science and Computer Engineering (CESCE) | 7 | 14 |
| 1137 | Q4 | International Conference on Communications and Information Technology | 7 | 14 |
| 1137 | Q4 | International Conference on Computer Graphics, Imaging and Visualization, CGIV | 7 | 14 |
| 1137 | Q4 | International Conference on Intelligent Computing and Cognitive Informatics | 7 | 14 |
| 1137 | Q4 | Italian Conference on Computational Logic (CILC) | 7 | 14 |
| 1143 | Q4 | International Conference on Education and Management Technology (ICEMT) | 7 | 13 |
| 1143 | Q4 | International Conference on Network Computing and Information Security | 7 | 13 |
| 1143 | Q4 | International Work-Conference on the Interplay Between Natural and Artificial Computation | 7 | 13 |
| 1143 | Q4 | Symposium on Virtual and Augmented Reality | 7 | 13 |
| 1147 | Q4 | IEEE International Conference on Signal and Image Processing Applications | 7 | 12 |
| 1147 | Q4 | IEEE International Symposium for Design and Technology in Electronic Packaging (SIITME) | 7 | 12 |
| 1147 | Q4 | International Conference on Optical Internet | 7 | 12 |
| 1147 | Q4 | International Workshop on Pattern Recognition in Neuroimaging | 7 | 12 |
| 1147 | Q4 | Malaysian Conference in Software Engineering (MySEC) | 7 | 12 |
| 1147 | Q4 | Pervasive and Embedded Computing and Communication Systems (PECCS) | 7 | 12 |
| 1153 | Q4 | Asia-Pacific Conference on Wearable Computing Systems | 7 | 11 |
| 1153 | Q4 | CSI International Symposium on Artificial Intelligence and Signal Processing | 7 | 11 |
| 1153 | Q4 | IEEE Symposium on Communications and Vehicular Technology in the Benelux | 7 | 11 |
| 1153 | Q4 | International Conference on E-Business and Information System Security, EBISS | 7 | 11 |
| 1153 | Q4 | International Conference on Electronic Visualisation and the Arts | 7 | 11 |
| 1153 | Q4 | International Conference on Informatics and Systems | 7 | 11 |
| 1153 | Q4 | International Conference on Software Engineering & Computer Systems | 7 | 11 |
| 1153 | Q4 | International Semiconductor Device Research Symposium | 7 | 11 |
| 1153 | Q4 | International Workshop on Computational Electronics | 7 | 11 |
| 1153 | Q4 | World Congress on Software Engineering, WCSE | 7 | 11 |
| 1163 | Q4 | Artificial Intelligence: Methodology, Systems, Applications (AIMSA) | 7 | 10 |
| 1163 | Q4 | Catalan Conference on AI | 7 | 10 |
| 1163 | Q4 | Electric Power Quality and Supply Reliability Conference (PQ) | 7 | 10 |
| 1163 | Q4 | IEEE International Conference on Microwaves, Communications, Antennas and Electronics Systems | 7 | 10 |
| 1163 | Q4 | IEEE Wireless Power Transfer Conference | 7 | 10 |
| 1163 | Q4 | International Conference on Multimedia and Signal Processing | 7 | 10 |
| 1163 | Q4 | International Conference on Optoelectronics and Image Processing | 7 | 10 |
| 1163 | Q4 | International Conference on Recent Advancements in Electrical, Electronics and Control Engineering | 7 | 10 |
| 1163 | Q4 | International Conference on Research and Innovation in Information Systems | 7 | 10 |
| 1163 | Q4 | International Conference on Signals and Electronic Systems | 7 | 10 |
| 1163 | Q4 | International Conference on Simulation of Semiconductor Processes and Devices | 7 | 10 |
| 1163 | Q4 | International Conference on Soft Computing for Problem Solving | 7 | 10 |
| 1163 | Q4 | International Microsystems Packaging Assembly and Circuits Technology Conference | 7 | 10 |
| 1163 | Q4 | International Symposium on Advanced Control of Industrial Processes | 7 | 10 |
| 1163 | Q4 | International Symposium on Computational Intelligence and Design | 7 | 10 |
| 1163 | Q4 | International Symposium on Intelligence Information Processing and Trusted Computing (IPTC) | 7 | 10 |
| 1163 | Q4 | International Workshop on Signal Design and Its Applications in Communications | 7 | 10 |
| 1163 | Q4 | IPCC IEEE International Professional Communication Conference | 7 | 10 |
| 1163 | Q4 | Workshop on Integrated Nonlinear Microwave and Millimetre-Wave Circuits | 7 | 10 |
| 1163 | Q4 | WSEAS International Conference on Applied computer science | 7 | 10 |
| 1183 | Q4 | Asia Pacific Conference on Postgraduate Research in Microelectronics & Electronics PrimeAsia | 7 | 9 |
| 1183 | Q4 | Chinese Conference on Biometric Recognition | 7 | 9 |
| 1183 | Q4 | Cross Strait Quad-Regional Radio Science and Wireless Technology Conference (CSQRWC) | 7 | 9 |
| 1183 | Q4 | Electronics, Robotics and Automotive Mechanics Conference | 7 | 9 |
| 1183 | Q4 | IEEE International Conference on Networks | 7 | 9 |
| 1183 | Q4 | IEEE International Conference on Semiconductor Electronics | 7 | 9 |
| 1183 | Q4 | IEEE International Conference on Spatial Data Mining and Geographical Knowledge Services | 7 | 9 |
| 1183 | Q4 | IEEE International Conference on Technology for Education | 7 | 9 |
| 1183 | Q4 | International Asia-Pacific Conference on Synthetic Aperture Radar | 7 | 9 |
| 1183 | Q4 | International Conference on Advanced Mechatronic Systems | 7 | 9 |
| 1183 | Q4 | International Conference on Auditory-Visual Speech Processing (AVSP) | 7 | 9 |
| 1183 | Q4 | International Conference on Devices, Circuits and Systems | 7 | 9 |
| 1183 | Q4 | International Conference on Emerging Trends in Networks and Computer Communications (ETNCC) | 7 | 9 |
| 1183 | Q4 | International Conference on ICT and Knowledge Engineering (ICT & Knowledge Engineering) | 7 | 9 |
| 1183 | Q4 | International Conference on Information Technology and Multimedia | 7 | 9 |
| 1183 | Q4 | International Conference on Internet Multimedia Computing and Service | 7 | 9 |
| 1183 | Q4 | International Conference on Numerical Simulation of Optoelectronic Devices | 7 | 9 |
| 1183 | Q4 | International Conference on Power Electronics Systems and Applications | 7 | 9 |
| 1183 | Q4 | International Symposium on Chinese Spoken Language Processing (ISCSLP) | 7 | 9 |
| 1183 | Q4 | International Symposium on Systems and Control in Aerospace and Astronautics | 7 | 9 |
| 1183 | Q4 | Italian Research Conference on Digital Libraries (IRCDL) | 7 | 9 |
| 1183 | Q4 | Mobile, Ubiquitous, and Intelligent Computing (MUSIC) | 7 | 9 |
| 1183 | Q4 | WRI Global Congress on Intelligent Systems, GCIS | 7 | 9 |
| 1183 | Q4 | WSEAS International Conference on Mathematical and computational methods in science and engineering | 7 | 9 |
| 1207 | Q4 | Electrical Insulation Conference (EIC) | 7 | 8 |
| 1207 | Q4 | IEEE CIE International Conference on Radar | 7 | 8 |
| 1207 | Q4 | IEEE Conference on Computational Intelligence for Financial Engineering & Economics (CIFEr) | 7 | 8 |
| 1207 | Q4 | International Conference on Advances in Computing and Information Technology | 7 | 8 |
| 1207 | Q4 | International Conference on Business Management and Electronic Information (BMEI) | 7 | 8 |
| 1207 | Q4 | International Conference on Computer & Information Science | 7 | 8 |
| 1207 | Q4 | International Conference on Intelligent Human Computer Interaction (IHCI) | 7 | 8 |
| 1207 | Q4 | International Conference on Modern Problems of Radio Engineering Telecommunications and Computer Science | 7 | 8 |
| 1207 | Q4 | International Symposium on Knowledge Acquisition and Modeling | 7 | 8 |
| 1207 | Q4 | Microwaves, Radar and Remote Sensing Symposium (MRRS) | 7 | 8 |
| 1207 | Q4 | NIP & Digital Fabrication Conference | 7 | 8 |
| 1218 | Q4 | IEEE Global Conference on Consumer Electronics (GCCE) | 7 | 7 |
| 1218 | Q4 | International Conference on Artificial Intelligence and Education (ICAIE) | 7 | 7 |
| 1218 | Q4 | International Conference on Electric Power and Energy Conversion Systems | 7 | 7 |
| 1218 | Q4 | International Conference on Electrical and Electronics Engineering | 7 | 7 |
| 1222 | Q4 | International Workshop on Database Technology and Applications (DBTA) | 6 | 18 |
| 1223 | Q4 | International Workshop on Software Measurement (IWSM) | 6 | 12 |
| 1224 | Q4 | Bioinformatics & Computational Biology (BIOCOMP) | 6 | 11 |
| 1224 | Q4 | IEEE International Conference on Virtual Environments, Human-Computer Interfaces and Measurements Systems | 6 | 11 |
| 1224 | Q4 | International Conference on Computing, Management and Telecommunications (ComManTel) | 6 | 11 |
| 1224 | Q4 | International Conference on Security Protocols | 6 | 11 |
| 1224 | Q4 | Specialist Meeting on Microwave Radiometry and Remote Sensing of the Environment (MicroRad) | 6 | 11 |
| 1229 | Q4 | Chinese Conference on Pattern Recognition | 6 | 10 |
| 1229 | Q4 | IEEE International Conference on Microwave Technology & Computational Electromagnetics | 6 | 10 |
| 1229 | Q4 | IEEE International Conference on Space Science and Communication (IconSpace) | 6 | 10 |
| 1229 | Q4 | IEEE Symposium on Web Society (SWS) | 6 | 10 |
| 1229 | Q4 | IITA International Conference on Geoscience and Remote Sensing | 6 | 10 |
| 1229 | Q4 | International Conference on Communication Systems, Networks and Applications (ICCSNA) | 6 | 10 |
| 1229 | Q4 | International Conference on Optical Communications and Networks | 6 | 10 |
| 1236 | Q4 | Computer Science and Electronic Engineering Conference (CEEC) | 6 | 9 |
| 1236 | Q4 | IEEE Conference on Evolving and Adaptive Intelligent Systems | 6 | 9 |
| 1236 | Q4 | IEEE Youth Conference on Information, Computing and Telecommunication YC-ICT | 6 | 9 |
| 1236 | Q4 | International Conference on Advanced Computing, Networking and Security | 6 | 9 |
| 1236 | Q4 | International Conference on Communications, Signal Processing, and their Applications | 6 | 9 |
| 1236 | Q4 | International Conference on Computer Science and Information Processing | 6 | 9 |
| 1236 | Q4 | International Conference on Control, Automation and Systems Engineering (CASE) | 6 | 9 |
| 1236 | Q4 | International Conference on Information Science and Digital Content Technology | 6 | 9 |
| 1236 | Q4 | International Conference on Instrumentation, Measurement, Computer, Communication and Control | 6 | 9 |
| 1236 | Q4 | International Conference on Sensor Networks (SENSORNETS) | 6 | 9 |
| 1236 | Q4 | International Conference on Telecommunications in Modern Satellite, Cable and Broadcasting Services | 6 | 9 |
| 1236 | Q4 | International Symposium on Integrated Circuits | 6 | 9 |
| 1236 | Q4 | International Symposium on Intelligent Informatics (ISI) | 6 | 9 |
| 1236 | Q4 | International Symposium on Visual Information Communication and Interaction | 6 | 9 |
| 1236 | Q4 | International Technology, Education and Development Conference | 6 | 9 |
| 1236 | Q4 | Spanish Conference on Electron Devices CDE | 6 | 9 |
| 1252 | Q4 | IEEE AESS European Conference on Satellite Telecommunications (ESTEL) | 6 | 8 |
| 1252 | Q4 | IEEE International Conference on Advanced Computational Intelligence (ICACI) | 6 | 8 |
| 1252 | Q4 | IEEE International Conference on Healthcare Informatics | 6 | 8 |
| 1252 | Q4 | IEEE Jordan Conference on Applied Electrical Engineering and Computing Technologies | 6 | 8 |
| 1252 | Q4 | IMOC SBMO/IEEE MTT-S International Microwave and Optoelectronics Conference | 6 | 8 |
| 1252 | Q4 | International Conference on Circuit, Power and Computing Technologies | 6 | 8 |
| 1252 | Q4 | International Conference on Data Technologies and Applications (DATA) | 6 | 8 |
| 1252 | Q4 | International Conference on Education and e-Learning Innovations | 6 | 8 |
| 1252 | Q4 | International Conference on Educational and Network Technology | 6 | 8 |
| 1252 | Q4 | International Conference on Flexible Query Answering Systems (FQAS) | 6 | 8 |
| 1252 | Q4 | International Conference on Fluid Power and Mechatronics (FPM) | 6 | 8 |
| 1252 | Q4 | International Conference on Information Communication and Embedded Systems | 6 | 8 |
| 1252 | Q4 | International Conference on Intelligent Systems and Control | 6 | 8 |
| 1252 | Q4 | International Conference on Power, Signals, Controls and Computation | 6 | 8 |
| 1252 | Q4 | International ISC Conference on Information Security and Cryptology | 6 | 8 |
| 1252 | Q4 | International Kharkov Symposium on Physics and Engineering of Microwaves, Millimeter and Submillimeter Waves and Workshop on Terahertz Technologies | 6 | 8 |
| 1252 | Q4 | International Symposium on Antennas and Propagation | 6 | 8 |
| 1252 | Q4 | International Symposium on Electrical and Electronics Engineering | 6 | 8 |
| 1252 | Q4 | International Workshop on Combinatorial Image Analysis | 6 | 8 |
| 1252 | Q4 | Proceedings of the First International Conference on Bioinformatics (BIOINFORMATICS) | 6 | 8 |
| 1272 | Q4 | Annual International Conference on Electronic Commerce | 6 | 7 |
| 1272 | Q4 | European Microelectronics Packaging Conference | 6 | 7 |
| 1272 | Q4 | IEEE Electrical Design of Advanced Packaging & Systems Symposium (EDAPS) | 6 | 7 |
| 1272 | Q4 | IEEE International Symposium on Next-Generation Electronics | 6 | 7 |
| 1272 | Q4 | International Conference and Seminar on Micro/Nanotechnologies and Electron Devices | 6 | 7 |
| 1272 | Q4 | International Conference on Asian Language Processing | 6 | 7 |
| 1272 | Q4 | International Conference on Computer Aided Systems Theory | 6 | 7 |
| 1272 | Q4 | International Conference on E-Business and Telecommunication Networks | 6 | 7 |
| 1272 | Q4 | International Conference on Electronic Measurement & Instruments | 6 | 7 |
| 1272 | Q4 | International Conference on Industrial Control and Electronics Engineering | 6 | 7 |
| 1272 | Q4 | International Conference on Model-Driven Engineering and Software Development | 6 | 7 |
| 1272 | Q4 | International Conference on Robot, Vision and Signal Processing | 6 | 7 |
| 1272 | Q4 | International Conference on Signal Processing and Integrated Networks | 6 | 7 |
| 1272 | Q4 | International Conference on Technological Advances in Electrical, Electronics and Computer Engineering | 6 | 7 |
| 1272 | Q4 | International Silicon-Germanium Technology and Device Meeting | 6 | 7 |
| 1272 | Q4 | International Symposium on Instrumentation and Measurement, Sensor Network and Automation | 6 | 7 |
| 1272 | Q4 | International Workshop on Advanced Ground Penetrating Radar | 6 | 7 |
| 1272 | Q4 | Korea-Japan Joint Workshop on Frontiers of Computer Vision | 6 | 7 |
| 1272 | Q4 | MATEC Web of Conferences | 6 | 7 |
| 1272 | Q4 | Mediterranean Conference on Information Systems (MCIS) | 6 | 7 |
| 1272 | Q4 | Signal Processing and Multimedia Applications (SIGMAP) | 6 | 7 |
| 1293 | Q4 | Anais do Workshop de Informática na Escola | 6 | 6 |
| 1293 | Q4 | Electric Vehicle Symposium and Exhibition (EVS27) | 6 | 6 |
| 1293 | Q4 | IEEE International Workshop on Genomic Signal Processing and Statistics | 6 | 6 |
| 1293 | Q4 | International Conference on Transportation, Mechanical, and Electrical Engineering | 6 | 6 |
| 1293 | Q4 | SHS Web of Conferences | 6 | 6 |
| 1298 | Q4 | Spring/Summer Young Researchers' Colloquium on Software Engineering | 5 | 22 |
| 1299 | Q4 | Conference on Optoelectronic and Microelectronic Materials and Devices (COMMAD) | 5 | 20 |
| 1300 | Q4 | Brazilian Symposium on Multimedia and the Web | 5 | 12 |
| 1301 | Q4 | IEEE IAS Electrical Safety Workshop | 5 | 10 |
| 1301 | Q4 | International Siberian Conference on Control and Communications (SIBCON) | 5 | 10 |
| 1303 | Q4 | Brazilian Symposium on Computing System Engineering | 5 | 9 |
| 1303 | Q4 | Conference on Advances in Communication and Control Systems | 5 | 9 |
| 1303 | Q4 | Indian International Conference on Artificial Intelligence | 5 | 9 |
| 1303 | Q4 | International Conference on Image Processing, Computer Vision, & Pattern Recognition | 5 | 9 |
| 1303 | Q4 | International Conference on Laser and Fiber-Optical Networks Modeling | 5 | 9 |
| 1303 | Q4 | International Conference on Ubiquitous Computing and Ambient Intelligence | 5 | 9 |
| 1303 | Q4 | International Conference on Wireless Communication and Sensor Networks | 5 | 9 |
| 1303 | Q4 | International Symposium on Computers in Education | 5 | 9 |
| 1303 | Q4 | International Symposium on Mechatronics and its Applications | 5 | 9 |
| 1303 | Q4 | Review of Business Information Systems (RBIS) | 5 | 9 |
| 1313 | Q4 | IEEE China Summit & International Conference on Signal and Information Processing (ChinaSIP) | 5 | 8 |
| 1313 | Q4 | IEEE International Conference on Computer-Aided Design and Computer Graphics | 5 | 8 |
| 1313 | Q4 | International Conference on Advanced Intelligence and Awarenss Internet | 5 | 8 |
| 1313 | Q4 | International Conference on Communication, Electronics and Automation Engineering | 5 | 8 |
| 1313 | Q4 | International Conference on Control Engineering and Communication Technology | 5 | 8 |
| 1313 | Q4 | International Conference on Digital Home | 5 | 8 |
| 1313 | Q4 | International Conference on Emerging Trends in Computing, Communication and Nanotechnology | 5 | 8 |
| 1313 | Q4 | International Symposium on Advanced Packaging Materials (APM) | 5 | 8 |
| 1313 | Q4 | International Symposium on Biometrics and Security Technologies | 5 | 8 |
| 1313 | Q4 | International Symposium on Information Engineering and Electronic Commerce | 5 | 8 |
| 1313 | Q4 | National Conference on Computer Vision, Pattern Recognition, Image Processing and Graphics | 5 | 8 |
| 1324 | Q4 | BIO Web of Conferences | 5 | 7 |
| 1324 | Q4 | Electrical Overstress/Electrostatic Discharge Symposium (EOS/ESD) | 5 | 7 |
| 1324 | Q4 | IEEE International Conference on Emerging eLearning Technologies & Applications (ICETA) | 5 | 7 |
| 1324 | Q4 | IEEE International Wireless Symposium | 5 | 7 |
| 1324 | Q4 | IEEE Nanotechnology Materials and Devices Conference | 5 | 7 |
| 1324 | Q4 | IEEE/CPMT International Electronic Manufacturing Technology Symposium (IEMT) | 5 | 7 |
| 1324 | Q4 | International Conference on Advanced Computer Science and Information Systems | 5 | 7 |
| 1324 | Q4 | International Conference on Communications, Devices and Intelligent Systems | 5 | 7 |
| 1324 | Q4 | International Conference on Computational Intelligence and Natural Computing, CINC | 5 | 7 |
| 1324 | Q4 | International Conference on Current Research Information Systems (CRIS) | 5 | 7 |
| 1324 | Q4 | International Conference on Electric Power Equipment-Switching Technology | 5 | 7 |
| 1324 | Q4 | International Conference on Electric Technology and Civil Engineering (ICETCE) | 5 | 7 |
| 1324 | Q4 | International Conference on Frontiers in Computer Education (ICFCE) | 5 | 7 |
| 1324 | Q4 | International Conference on IT Convergence and Security (ICITCS) | 5 | 7 |
| 1324 | Q4 | International Conference on Logistics, Informatics and Service Science | 5 | 7 |
| 1324 | Q4 | International Conference on Remote Sensing, Environment and Transportation Engineering | 5 | 7 |
| 1324 | Q4 | International Conference on Systems Biology | 5 | 7 |
| 1324 | Q4 | International Symposium on Image and Data Fusion (ISIDF) | 5 | 7 |
| 1324 | Q4 | International Work-Conference on Bioinformatics and Biomedical Engineering (IWBBIO) | 5 | 7 |
| 1324 | Q4 | Joint International Conference on Supercomputing in Nuclear Applications | 5 | 7 |
| 1344 | Q4 | Europe-Asia Congress on Mechatronics (MECATRONICS) | 5 | 6 |
| 1344 | Q4 | IEEE Conference on Business Informatics | 5 | 6 |
| 1344 | Q4 | IEEE International Conference on Computer-Aided Industrial Design & Conceptual Design | 5 | 6 |
| 1344 | Q4 | IEEE International Conference on Electronics, Computing and Communication Technologies | 5 | 6 |
| 1344 | Q4 | IEEE Symposium on Computer Applications and Industrial Electronics | 5 | 6 |
| 1344 | Q4 | IEEE Symposium on Electrical & Electronics Engineering | 5 | 6 |
| 1344 | Q4 | International Aegean Conference on Electrical Machines and Power Electronics | 5 | 6 |
| 1344 | Q4 | International Conference on Advanced Computing, Networking and Informatics (ICACNI) | 5 | 6 |
| 1344 | Q4 | International Conference on Advances in Computing | 5 | 6 |
| 1344 | Q4 | International Conference on e-Learning and e-Technologies in Education | 5 | 6 |
| 1344 | Q4 | International Conference on Information, Intelligence, Systems and Applications | 5 | 6 |
| 1344 | Q4 | International Conference on Issues and Challenges in Intelligent Computing Techniques | 5 | 6 |
| 1344 | Q4 | International Conference on Model and data engineering | 5 | 6 |
| 1344 | Q4 | International Conference on Operations Reserach and Enterprise Systems (ICORES) | 5 | 6 |
| 1344 | Q4 | International Conference on Practical Applications of Computational Biology & Bioinformatics | 5 | 6 |
| 1344 | Q4 | International Electronic Conference on Synthetic Organic Chemistry | 5 | 6 |
| 1344 | Q4 | International Symposium on Electrical Insulating Materials | 5 | 6 |
| 1344 | Q4 | International Symposium on Microwave, Antenna, Propagation and EMC Technologies for Wireless Communications | 5 | 6 |
| 1344 | Q4 | International Visual Informatics Conference | 5 | 6 |
| 1344 | Q4 | Multi-disciplinary Trends in Artificial Intelligence (MIWAI) | 5 | 6 |
| 1344 | Q4 | RSI/ISM International Conference on Robotics and Mechatronics | 5 | 6 |
| 1344 | Q4 | Sino-foreign-interchange Conference on Intelligent Science and Intelligent Data Engineering | 5 | 6 |
| 1344 | Q4 | Symposium on Piezoelectricity, Acoustic Waves, and Device Applications (SPAWDA) | 5 | 6 |
| 1344 | Q4 | Web Information Systems and Applications Conference (WISA) | 5 | 6 |
| 1368 | Q4 | Digital Heritage International Congress (DigitalHeritage) | 5 | 5 |
| 1368 | Q4 | IIAI International Conference on Advanced Applied Informatics | 5 | 5 |
| 1368 | Q4 | International Symposium on Bioelectronics and Bioinformatics | 5 | 5 |
| 1368 | Q4 | International Symposium on VLSI Design and Test | 5 | 5 |
| 1368 | Q4 | Power Systems Computation Conference (PSCC) | 5 | 5 |
| 1368 | Q4 | Studia i Materialy Polskiego Stowarzyszenia Zarzadzania Wiedza/Studies & Proceedings Polish Association for Knowledge Management | 5 | 5 |
| 1374 | Q4 | E3S Web of Conferences | 4 | 13 |
| 1374 | Q4 | Technologies Applied to Electronics Teaching (TAEE) | 4 | 13 |
| 1376 | Q4 | Frontiers of Intelligent Computing: Theory and Applications (FICTA) | 4 | 12 |
| 1377 | Q4 | International Conference on Wireless Technologies for Humanitarian Relief | 4 | 9 |
| 1378 | Q4 | International Conference on Computing, Measurement, Control and Sensor Network | 4 | 8 |
| 1379 | Q4 | IEEE Symposium on Robotics and Applications | 4 | 7 |
| 1379 | Q4 | International Conference on Internet Computing in Science and Engineering | 4 | 7 |
| 1379 | Q4 | International Conference on Speech and Computer | 4 | 7 |
| 1379 | Q4 | International Conference on Virtual Reality and Visualization | 4 | 7 |
| 1383 | Q4 | International Conference on Communications, Computers and Applications | 4 | 6 |
| 1383 | Q4 | International Conference on Computer Applications Technology | 4 | 6 |
| 1383 | Q4 | International Conference on Instrumentation, Communications, Information Technology, and Biomedical Engineering | 4 | 6 |
| 1383 | Q4 | International Conference on Sciences of Electronics, Technologies of Information and Telecommunications | 4 | 6 |
| 1383 | Q4 | International Conference on Teaching and Learning in Computing and Engineering | 4 | 6 |
| 1383 | Q4 | International Symposium on Semiconductor Manufacturing | 4 | 6 |
| 1383 | Q4 | National Conference on Electrical, Electronics and Computer Engineering | 4 | 6 |
| 1383 | Q4 | Subthreshold Microelectronics Technology Unified SOI Conference | 4 | 6 |
| 1391 | Q4 | Computer und Roboterassistierte Chirugie (CURAC) | 4 | 5 |
| 1391 | Q4 | Conference on Software Engineering Education | 4 | 5 |
| 1391 | Q4 | Conferencia Latinoamericana En Informatica (CLEI) | 4 | 5 |
| 1391 | Q4 | Euro American Conference on Telematics and Information Systems | 4 | 5 |
| 1391 | Q4 | Iberian Robotics Conference | 4 | 5 |
| 1391 | Q4 | International Caribbean Conference on Devices, Circuits and Systems, ICCDCS | 4 | 5 |
| 1391 | Q4 | International Conference on Actual Problems of Electronic Instrument Engineering | 4 | 5 |
| 1391 | Q4 | International Conference on Advances in Communication, Network, and Computing | 4 | 5 |
| 1391 | Q4 | International Conference on Antenna Theory and Techniques | 4 | 5 |
| 1391 | Q4 | International Conference on Applied Superconductivity and Electromagnetic Devices, ASEMD | 4 | 5 |
| 1391 | Q4 | International Conference on Computational Science and Computational Intelligence | 4 | 5 |
| 1391 | Q4 | International Conference on Computers and Devices for Communication | 4 | 5 |
| 1391 | Q4 | International Conference on Computing, Electrical and Electronics Engineering | 4 | 5 |
| 1391 | Q4 | International Conference on Control, Decision and Information Technologies | 4 | 5 |
| 1391 | Q4 | International Conference on Electrical Engineering and Software Applications | 4 | 5 |
| 1391 | Q4 | International Conference on Information and Software Technologies (ICIST) | 4 | 5 |
| 1391 | Q4 | International Conference on Pervasive Computing and the Networked World | 4 | 5 |
| 1391 | Q4 | International Conference on Thermal, Mechanical and Multi-Physics Simulation and Experiments in Microelectronics and Microsystems (EuroSimE) | 4 | 5 |
| 1391 | Q4 | International Congress on Nursing Informatics | 4 | 5 |
| 1391 | Q4 | International Symposium on Computing and Networking | 4 | 5 |
| 1391 | Q4 | International Workshop on Active-Matrix Flatpanel Displays and Devices | 4 | 5 |
| 1412 | Q4 | Anais dos Workshops do Congresso Brasileiro de Informática na Educação | 4 | 4 |
| 1412 | Q4 | Asia-Pacific Conference on Antennas and Propagation | 4 | 4 |
| 1412 | Q4 | Asia-Pacific Conference on Information Network and Digital Content Security | 4 | 4 |
| 1412 | Q4 | Asia-Pacific Conference on Information Theory | 4 | 4 |
| 1412 | Q4 | Brazilian Conference on Intelligent Systems | 4 | 4 |
| 1412 | Q4 | IEEE International Scientific Conference Electronics and Nanotechnology (ELNANO) | 4 | 4 |
| 1412 | Q4 | IEEE Malaysia International Conference on Communications (MICC) | 4 | 4 |
| 1412 | Q4 | International Conference on Advanced Semiconductor Devices & Microsystems (ASDAM) | 4 | 4 |
| 1412 | Q4 | International Conference on Applied Informatics and Communication | 4 | 4 |
| 1412 | Q4 | International Conference on Electronics, Computer and Computation | 4 | 4 |
| 1412 | Q4 | International Conference on Emerging Technology Trends in Electronics, Communication and Networking | 4 | 4 |
| 1412 | Q4 | International Conference on Fuzzy Theory and Its Applications | 4 | 4 |
| 1412 | Q4 | International Conference on Human Computer Interaction | 4 | 4 |
| 1412 | Q4 | International Conference on Optoelectronics and Microelectronics | 4 | 4 |
| 1412 | Q4 | International Conference on Systems and Control | 4 | 4 |
| 1427 | Q4 | International Conference on Green Computing Communication and Electrical Engineering | 3 | 28 |
| 1428 | Q4 | International Conference on Engineering and Technology Education | 3 | 12 |
| 1429 | Q4 | Conference on Computational Linguistics and Speech Processing (ROCLING) | 3 | 8 |
| 1430 | Q4 | International Conference on Next Generation Networks and Services | 3 | 7 |
| 1431 | Q4 | International Conference in Electrics, Communication and Automatic Control | 3 | 6 |
| 1432 | Q4 | IET International Conference on Information Science and Control Engineering | 3 | 5 |
| 1432 | Q4 | International Conference on Advanced Computer Science and Electronics Information | 3 | 5 |
| 1432 | Q4 | International Conference on Electrical Information and Communication Technology | 3 | 5 |
| 1432 | Q4 | International Conference on Engineering and Computer Education | 3 | 5 |
| 1432 | Q4 | International Conference on Information Technology and Software Engineering | 3 | 5 |
| 1432 | Q4 | International Future Energy Electronics Conference (IFEEC) | 3 | 5 |
| 1432 | Q4 | International Workshop on Earth Observation and Remote Sensing Applications | 3 | 5 |
| 1439 | Q4 | IEEE International Conference on Oxide Materials for Electronic Engineering | 3 | 4 |
| 1439 | Q4 | IEEE International Meeting for Future of Electron Devices, Kansai (IMFEDK) | 3 | 4 |
| 1439 | Q4 | IEEE International Microwave and RF Conference | 3 | 4 |
| 1439 | Q4 | International Conference on Cloud Computing and Big Data | 3 | 4 |
| 1439 | Q4 | International Conference on Computer Technologies in Physical and Engineering Applications | 3 | 4 |
| 1439 | Q4 | International Conference on Electrical Engineering and Information & Communication Technology | 3 | 4 |
| 1439 | Q4 | International Conference on Future Computer Science and Education (ICFCSE) | 3 | 4 |
| 1439 | Q4 | International Conference on Mining Intelligence and Knowledge Exploration | 3 | 4 |
| 1439 | Q4 | International Conference on Wavelet Active Media Technology and Information Processing | 3 | 4 |
| 1439 | Q4 | International Conference on Wireless Communications, Vehicular Technology, Information Theory and Aerospace & Electronic Systems | 3 | 4 |
| 1439 | Q4 | International Conference Problems of Cybernetics and Informatics | 3 | 4 |
| 1439 | Q4 | International Congress on Advanced Electromagnetic Materials in Microwaves and Optics | 3 | 4 |
| 1439 | Q4 | International Crimean Conference on Microwave and Telecommunication Technology | 3 | 4 |
| 1439 | Q4 | International Workshop on Microwave and Millimeter Wave Circuits and System Technology | 3 | 4 |
| 1439 | Q4 | Symposium of Image, Signal Processing, and Artificial Vision (STSIVA) | 3 | 4 |
| 1454 | Q4 | Anais do Congresso Nacional Universidade, EAD e Software Livre | 3 | 3 |
| 1454 | Q4 | International Conference on Electronics Packaging | 3 | 3 |
| 1454 | Q4 | International Conference on Information Systems for Crisis Response and Management (ISCRAM) | 3 | 3 |
| 1454 | Q4 | International Conference on Information Technology and Electrical Engineering | 3 | 3 |
| 1454 | Q4 | International Conference on Information, Business and Education Technology | 3 | 3 |
| 1454 | Q4 | Iranian Conference on Fuzzy Systems | 3 | 3 |
| 1460 | Q4 | International Conference on Convergence Computer Technology | 2 | 7 |
| 1461 | Q4 | Symposium on Microelectronics Technology and Devices (SBMicro) | 2 | 5 |
| 1461 | Q4 | World Congress on Computer Applications and Information Systems | 2 | 5 |
| 1463 | Q4 | International Conference on Sciences and Techniques of Automatic Control and Computer Engineering | 2 | 4 |
| 1463 | Q4 | Electronic International Interdisciplinary in EIIC Conference | 2 | 3 |
| 1463 | Q4 | IEEE International Autumn Meeting on Power, Electronics and Computing | 2 | 3 |
| 1463 | Q4 | International Conference on Artificial Intelligence, Modelling and Simulation | 2 | 3 |
| 1463 | Q4 | International Conference on Cybernetics and Informatics | 2 | 3 |
| 1463 | Q4 | International Conference on Education and e-Learning | 2 | 3 |
| 1463 | Q4 | International Conference on Electric and Electronics | 2 | 3 |
| 1463 | Q4 | International Conference on Microwave and Photonics | 2 | 3 |
| 1463 | Q4 | International Conference on Mixed Design of Integrated Circuits and System, MIXDES | 2 | 3 |
| 1472 | Q4 | European Conference on Information Technology in Education and Society: A Critical Insight | 2 | 2 |
| 1472 | Q4 | IEEE International Conference on Big Data and Cloud Computing (BdCloud) | 2 | 2 |
| 1472 | Q4 | IEEE International Conference on MOOC, Innovation and Technology in Education | 2 | 2 |
| 1472 | Q4 | IEEE Workshop on Electronics, Computer and Applications | 2 | 2 |
| 1472 | Q4 | International Computer Conference on Wavelet Active Media Technology and Information Processing | 2 | 2 |
| 1472 | Q4 | International Conference on Actual Problems of Electron Devices Engineering | 2 | 2 |
| 1472 | Q4 | International Conference on Contemporary Computing and Informatics | 2 | 2 |
| 1472 | Q4 | International Workshop on Cloud Computing and Information Security | 2 | 2 |
| 1472 | Q4 | Telecommunications Symposium International | 2 | 2 |
| 1472 | Q4 | Virtual Reality Society of Japan, Annual 日本 バーチャル リアリティ 学会 大会 論文集 Conference | 2 | 2 |
| 1482 | Q4 | International Conference on Information Security and Intelligence Control | 1 | 4 |
| 1482 | Q4 | International Conference on Knowledge Management: Projects, Systems and Technologies | 1 | 4 |
| 1484 | Q4 | International Conference on Electrical Sciences and Technologies in Maghreb | 1 | 3 |
| 1485 | Q4 | Anais do Encontro Virtual de Documentação em Software Livre e Congresso Internacional de Linguagem e Tecnologia Online | 1 | 2 |
| 1485 | Q4 | IEEE Conference on Antenna Measurements & Applications | 1 | 2 |
| 1485 | Q4 | International Conference on Advances in Electronics, Computers and Communications | 1 | 2 |
| 1485 | Q4 | International Conference on Communications, Signal Processing, and Systems | 1 | 2 |
| 1485 | Q4 | International Conference on Education, Management and Computing Technology | 1 | 2 |
| 1485 | Q4 | International Conference on Mechatronics-Mechatronika | 1 | 2 |
| 1485 | Q4 | International Electrical Engineering Congress | 1 | 2 |
| 1485 | Q4 | the International Conference on Education Technology and Information System | 1 | 2 |
| 1493 | Q4 | Computational Intelligence on Power, Energy and Controls with their impact on Humanity (CIPECH), 2014 Innovative Applications of | 1 | 1 |
| 1493 | Q4 | IEEE International Conference on Consumer Electronics-Taiwan | 1 | 1 |
| 1493 | Q4 | International Conference on Computation of Power, Energy, Information and Communication | 1 | 1 |
| 1493 | Q4 | International Conference on Electronic Design | 1 | 1 |
| 1493 | Q4 | International Conference on Foreign Language Teaching and Applied Linguistics | 1 | 1 |
| 1493 | Q4 | International Conference on Mechanical Engineering, Automation and Control Systems | 1 | 1 |
| 1493 | Q4 | International Conference on Mechatronics, Control and Electronic Engineering | 1 | 1 |
| 1493 | Q4 | International Conference on Mechatronics, Electronic, Industrial and Control Engineering | 1 | 1 |

## ACKNOWLEDGEMENTS

Juan Manuel Ayllón has a FPI pre-doctoral scholarship for research (BES-2012- 054980) funded by the Spanish Ministry of Economy and Competitiveness. Alberto Martín-Martín holds a fellowship for the training of university teachers (FPU2013/05863), funded by the Spanish Ministry of Education, Culture and Sport.